\documentclass{mn2e}
\usepackage{graphicx}

\begin{document}

\title[Simultaneous p-mode and g-mode pulsation of HD 209295]
{Discovery and analysis of p-mode and g-mode oscillations in the A-type
primary of the eccentric binary HD 209295\thanks{Dedicated to the memory
of A. W. J. Cousins, discoverer of the variability of $\gamma$ Doradus}}
\author[G. Handler et al.]
       {G. Handler,$^{1}$ L. A. Balona,$^{1}$ R. R. Shobbrook,$^{2,3}$
	C. Koen,$^{1}$ A. Bruch,$^4$ \and E. Romero-Colmenero,$^1$
        A. A. Pamyatnykh,$^{5,6}$ B. Willems,$^{7}$ L. Eyer,$^{8,9}$\and
        D. J. James,$^{10,11,12,13}$ T. Maas$^{8}$
	\and \\
$^{1}$ South African Astronomical Observatory, P.O. Box 9, Observatory 7935,
South Africa\\
$^{2}$ P. O. Box 518, Coonabarabran, N.S.W 2357, Australia\\
$^{3}$ Research School of Astronomy and Astrophysics,
Australian National University, Weston Creek P.O., ACT 2611, Australia\\
$^4$ Lab\'oratorio Nacional de Astrof\'{\i}sica, Itajub\'a, Brazil\\
$^{5}$ Copernicus Astronomical Center, ul. Bartycka 18, 00-716 Warsaw,
Poland\\
$^6$ Institute of Astronomy, Russian Academy of Sciences,
Pyatnitskaya~48, 109017~Moscow, Russia\\
$^{7}$ Department of Physics and Astronomy, The Open University,
Walton Hall, Milton Keynes MK7 6AA, UK\\
$^{8}$ Instituut voor Sterrenkunde, Katholieke Universiteit Leuven,
B-3001 Leuven (Heverlee), Belgium\\
$^{9}$ Astrophysical Sciences Dept., Princeton University, Princeton,
New Jersey 08544, USA\\
$^{10}$ Observatoire de Gen\`{e}ve, Chemin des Maillettes 51, CH-1290
Sauverny, Switzerland\\
$^{11}$ Laboratoire d'Astrophysique, Observatoire de Grenoble,
Universit\'{e}
Joseph Fourier, B.P. 53, F-38041, Grenoble Cedex 9, France\\
$^{12}$ School of Physics \& Astronomy, University of St Andrews, North
Haugh, St Andrews, FIFE, KY16 9SS, United Kingdom\\
$^{13}$ 475 N. Charter Street, 5534 Sterling Hall, Madison, WI 53706-1582,
USA}

\date{Accepted 2001 nnnn nn.
      Received 2001 nnnn nn;
      in original form 2001 nnnn nn}

\maketitle

\begin{abstract} 

We have discovered both intermediate-order gravity mode and low-order
pressure mode pulsation in the same star, HD 209295. It is therefore both
a $\gamma$ Doradus and a $\delta$~Scuti star, which makes it the first
pulsating star to be a member of two classes.

The analysis of our 128 h of multi-site spectroscopic observations carried
out over two seasons reveals that the star is a single-lined spectroscopic
binary with an orbital period of 3.10575 $\pm$ 0.00010 d and an
eccentricity of 0.352 $\pm$ 0.011. Only weak pulsational signals are found
in both the radial velocity and line-profile variations, but we have
succeeded in showing that the two highest-amplitude $\gamma$ Doradus
pulsation modes are consistent with $\ell=1$ and $|m|=1$.

These two modes dominated our 280 h of BVI$_{\rm c}$ multi-site
photometry, also obtained over two seasons. We detected altogether ten
frequencies in the light variations, one in the $\delta$~Scuti regime and
nine in the $\gamma$ Doradus domain. Five of the $\gamma$ Doradus
frequencies are exact integer multiples of the orbital frequency. This
observation leads us to suspect they are tidally excited. Attempts to
identify modes from the multicolour photometry failed.

We performed model calculations and a stability analysis of the
pulsations. The frequency range in which $\delta$~Scuti modes are excited
agree well with observations. However, our models do not show excitation
of $\gamma$~Doradus pulsations, although the damping is smaller in the
observed range. We also investigated tidal excitation of $\gamma$~Doradus
modes. Some of the observed harmonics of the orbital period were found to
be unstable. The observed orbital harmonics which are stable in the models
can be understood as linear combinations of the unstable modes.

We could not detect the secondary component of the system in infrared
photometry, suggesting that it may not be a main-sequence star. Archival
data of this star shows that it has a strong ultraviolet excess, the origin
of which is not known. The orbit of the primary is consistent with a
secondary mass of $M>1.04 M_{\sun}$, which is indicative of a neutron star,
although a white dwarf companion is not ruled out.

\end{abstract}

\begin{keywords}
stars: variables: $\delta$~Sct -- stars: oscillations -- stars:
individual: HD 209295 -- stars: binaries: close -- stars: binaries:
spectroscopic -- stars: neutron
\end{keywords}

\section{Introduction}

Four different classes of multi-mode pulsating variables are found near the
intersection of the classical instability strip and the main sequence.
However, asteroseismology (probing the stellar interior through such
pulsations) of these stars has proven to be more difficult than expected.
Nonetheless, the example of helioseismology and its enormous reward in
terms of the physical understanding of the Sun's interior (e.g. see Gough
2000) has been a great motivation for continuing efforts in probing the
structure of these stars by means of their pulsational properties.

Many attempts to detect solar-type oscillations in stars have resulted in
inconclusive results, although success has been claimed in some instances
(e.g. Kjeldsen et al. 1995). Because of their extremely low amplitude, 
it is only recently (e.g. Bedding et al. 2002) that we are beginning to see 
detections which may be considered significant. Successful asteroseismology 
based on the analysis of solar-type oscillations now appears immanent.

The rapidly oscillating Ap stars (see Kurtz \& Martinez 2000 for a recent
comprehensive review) are high-radial order p-mode pulsators. Their
pulsation spectra resemble that of the Sun. In these stars, only a
few independent modes are observed. Moreover, the presence of a strong
magnetic field is expected to modify the frequencies and eigenfunctions of
the modes. Successful modelling of these pulsations does not seem possible
at this time. Nevertheless, some progress is currently being made (e.g.
Cunha \& Gough 2000).

The third group of pulsators in this region of the HR Diagram are the 
$\delta$~Scuti stars, which are low-radial order p- and probably
mixed-mode pulsators (see Breger \& Montgomery 2000). For some of these
stars, tens of pulsations modes have been detected and their frequencies
determined with high precision (see e.g. Handler et al. 2000).
Although mean densities, and even asteroseismological distances (Handler et
al. 1997) have been estimated, the problem of mode identification and
shortcomings in the stellar models are major difficulties (see e.g. Pamyatnykh 
et al. 1998 for a discussion).

The $\gamma$~Doradus stars are high-overtone gravity (g) mode pulsators (Kaye 
et al. 1999a). In these stars only a few modes are excited to observable 
amplitude amongst a dense forest of possible modes. As a result, accurate 
starting values of the basic stellar parameters are required to enable 
observed and calculated frequencies to be matched.

In principle, it is possible for $\delta$~Scuti and $\gamma$~Doradus
pulsations to co-exist in a star as they occupy overlapping regions in the
HR diagram. This offers new opportunities for successful asteroseismology,
as noted by Handler (1999a). In such stars it may be possible to use the
$\delta$~Scuti pulsations to place strong constraints on the stellar
parameters, easing the mode identification problem of the $\gamma$~Dor
pulsations. In this way it may be possible to obtain information about the
deep interior which is unobtainable for pure $\delta$~Sct stars.

Consequently, Handler \& Shobbrook (2002) searched for $\delta$~Scuti
pulsations in all known candidate $\gamma$~Doradus stars located within
the $\delta$~Scuti instability strip and accessible from intermediate
southern geographical latitudes. About one third of all non-Am and non-Ap
stars in the lower instability strip are indeed $\delta$~Scuti stars
(Breger 1975). Because Am and Ap stars are rare (if not absent) amongst
the $\gamma$~Doradus stars (Handler 1999a), detection of some ``hybrid''
stars may be expected if $\delta$~Scuti and $\gamma$~Doradus pulsations
are not mutually exclusive. In this paper we report the discovery and
analysis of such a star, HD 209295 = Sangam Mani.

This is a V=7.3 mag Southern ($\delta=-64\degr$) star of spectral type
A9/F0 V (Houk \& Cowley 1975). It was discovered as variable by the
{\it Hipparcos} mission (ESA 1997). Handler (1999a) performed a frequency
analysis of these data and detected four periods; we show the results in
Fig.~1 and Table 1. We are aware that a search for multiple periods in
{\it Hipparcos} photometry is difficult and can lead to spurious results
(Eyer \& Grenon 1998). However, the case of HD 209295 is so simple and
convincing that an error in the analysis is considered improbable. The
multi-periodicity leaves no doubt that HD 209295 is a {\it bona fide}
$\gamma$~Doradus star.

\begin{table}
\caption{Multi-frequency solution for the $\gamma$~Doradus variability of
HD~209295 from {\it Hipparcos} observations. Formal error estimates were
derived from Montgomery \& O'Donoghue (1999). The signal-to-noise ratio
was calculated following Breger et al. (1993); a S/N $\geq$ 4 corresponds
to a significant detection.}
\begin{center}
\begin{tabular}{cccc}
\hline
ID & Frequency & $H_{\rm p}$ Amplitude & S/N\\
 & (cycles/day) & (mmag)& \\
\hline
$f_1$ & 1.12957 $\pm$ 0.00002 & 37 $\pm$ 2 & 12\\
$f_2$ & 2.30222 $\pm$ 0.00003 & 31 $\pm$ 2 & 10 \\
$f_3$ & 2.57579 $\pm$ 0.00005 & 16 $\pm$ 2& 5\\
$f_4$ = $f_2 -f_1$ & 1.1726 $\pm$ 0.0001 & 12 $\pm$ 2 & 4\\
\hline
\end{tabular}
\end{center}
\end{table}

\begin{figure}
\includegraphics[width=90mm,viewport=46 10 384 478]{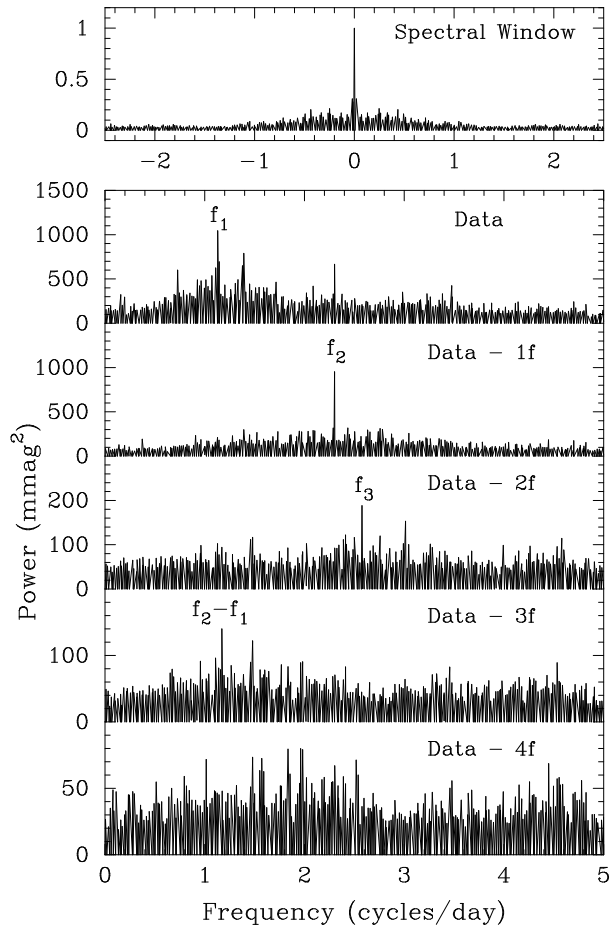}
\caption[]{Spectral window and amplitude spectra of {\it Hipparcos}
photometry of HD 209295 with consecutive prewhitening of detected
frequencies. The combination peak f$_2$-f$_1$ is only considered
significant because it occurs at a ``predicted'' frequency.}
\end{figure}

HD 209295 is considerably hotter than all other $\gamma$~Dor stars.
Handler (1999a) suggested that the published Str\"omgren $(b-y)$ of this
star (Twarog 1980) is too blue compared to $(B-V)$. Handler (1999b)
obtained new Str\"omgren photometry and confirmed the discrepancy (cf.
Table~2). Grenier et al. (1999) made two measurements of the radial
velocity of the star using an echelle spectrograph and the
cross-correlation technique. They did not comment on binarity or any other
peculiarity.

\begin{table}
\caption{Published uvby$\beta$ photometry of HD 209295. The differences
in the magnitudes and photometric indices are consistent with stellar
pulsation. When brighter the star it is bluer and the $c_1$ and 
H$_{\beta}$ indices are larger.}
\begin{center}
\begin{tabular}{cccccr}
\hline
V & $b-y$ & $m_1$ & $c_1$ & $\beta$ & Reference\\
\hline
7.32 & 0.149 & 0.182 & 0.819 & 2.781 & Twarog (1980) \\
7.29 & 0.139 & 0.185 & 0.840 & 2.821 & Handler (1999b)\\
\hline
\end{tabular}
\end{center}
\end{table}

Using the mean of the Str\"{o}mgren indices from Table 2 (which indicate
zero reddening), the effective temperature and gravity of HD~209295 can be
estimated. Kurucz's (1991) calibration yields $T_{\rm eff} = 7750 \pm
100$~K, $\log g = 4.10 \pm 0.05$.

The {\it Hipparcos} parallax of HD 209295 is $8.19 \pm 0.72$~mas. Using
$V_0 = 7.32 \pm 0.01$, $BC = -0.01 \pm 0.01$, the parallax gives $\log
L/L_{\sun} = 1.15 \pm 0.08$. With $\log T_{\rm eff} = 3.889$ as determined
above, $R/R_{\sun} = 2.08 \pm 0.09$ and, by means of evolutionary tracks
calculated using the Warsaw-New Jersey code (see e.g. Pamyatnykh et al.
1998), a mass of 1.84 $\pm 0.07 M_{\sun}$ is determined. We can then also
calculate a semi-independent second value for $\log g$, namely $4.07 \pm
0.05$, consistent with the previous value. HD 209295 is therefore the
hottest $\gamma$~Doradus star known to date; it lies in the middle of the
$\delta$~Scuti instability strip on the main sequence (Fig.~2).

\begin{figure}
\includegraphics[width=97mm,viewport=00 10 388 234]{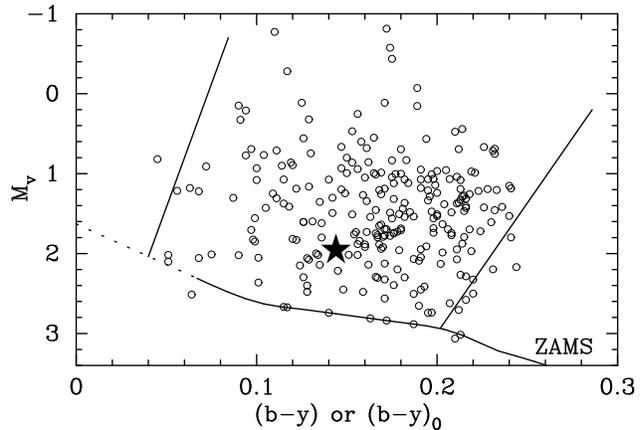}
\caption[]{The location of HD 209295 (star symbol) in the colour-magnitude
diagram compared to 252 known $\delta$~Sct stars (open circles) with
available Str\"omgren photometry from the catalogue of Rodriguez,
L\'opez-Gonz\'alez \& L\'opez de Coca (2000). The blue and red edges of
the $\delta$~Sct instability strip (Rodriguez \& Breger 2001) are indicated.}
\end{figure}

Because of its location in the colour-magnitude diagram, HD 209295 is a
good candidate $\delta$~Scuti star. Indeed, multi-periodic $\delta$~Scuti
pulsations were found by Handler \& Shobbrook (2002), thus placing HD
209295 as a member of two classes of pulsating star. Spectroscopic and
photometric follow-up observations were organized immediately after this
discovery, and a multi-site campaign was conducted in August 2000. We
report the results of these campaigns.

\section{Observations}

\subsection{Optical Photometry}

As comparison stars, we used HD 207802 (B9/B9.5V, $V=8.0$) and HD 209829 
(F3V, $V=7.7$) in our photoelectric photometry. Measurements were obtained at 
the 1.0-m, 0.75-m and 0.5-m telescopes of the Sutherland station of the South 
African Astronomical Observatory (SAAO), the 0.6-m telescope at Siding Spring 
Observatory (SSO) in Australia (Shobbrook 2000), and the 0.6-m Zeiss telescope 
at the Osservat\'orio do Pico dos Dias (LNA, Brazil). Observations were
conducted from October 1999 to October 2000. Most measurements were obtained 
in August and September 2000 in a co-ordinated multi-site effort.
A journal of the observations is given in Table 3.

\begin{table}
\caption[]{Journal of the photometric observations. The measurements
before HJD 2451600 were made in the Johnson BV system, whereas later
observations utilized Johnson-Cousins BVI$_{\rm c}$ except for one night
of BV observations marked with an asterisk.}
\begin{flushleft}
\begin{tabular}{lccr}
\hline
Telescope & Run start & Run length & Observer \\
 & HJD - 2450000 & (h) & \\
\hline
SAAO 0.75-m & 1464.335 & 3.31 & GH \\
SAAO 0.75-m & 1465.249 & 6.34 & GH \\
SSO 0.6-m & 1467.033 & 2.78 & RRS \\
SSO 0.6-m & 1468.916 & 6.05 & RRS \\
SSO 0.6-m & 1491.937 & 2.35 & RRS \\
SSO 0.6-m & 1498.918 & 2.35 & RRS \\
SAAO 0.5-m & 1502.282 & 1.85 & GH \\
SAAO 0.5-m & 1503.268 & 2.62 & GH \\
SAAO 0.5-m & 1505.267 & 2.57 & GH \\
SAAO 0.5-m & 1507.310 & 1.08 & GH \\
SAAO 0.5-m & 1509.295 & 1.63 & GH \\
SSO 0.6-m & 1510.924 & 1.85 & RRS \\
SAAO 0.5-m & 1511.264 & 2.40 & GH \\
SSO 0.6-m & 1512.933 & 2.18 & RRS \\
SSO 0.6-m & 1513.926 & 0.94 & RRS \\
SSO 0.6-m & 1516.930 & 2.04 & RRS \\
SSO 0.6-m & 1517.921 & 1.20 & RRS \\
SAAO 0.5-m & 1518.266 & 0.86 & LE \\
SAAO 0.5-m & 1519.283 & 0.48 & LE \\
SSO 0.6-m & 1524.930 & 1.63 & RRS \\
\hline
SSO 0.6-m & 1754.110 & 3.41 & RRS \\
SSO 0.6-m & 1754.962 & 5.45 & RRS \\
SAAO 0.75-m & 1758.343 & 3.46 & ERC \\
SSO 0.6-m & 1759.947 & 8.11 & RRS \\
SAAO 0.75-m & 1761.343 & 7.87 & ERC \\
SAAO 0.75-m & 1762.364 & 7.25 & ERC \\
SSO 0.6-m & 1762.944 & 7.75 & RRS \\
SAAO 0.75-m & 1763.337 & 8.11 & ERC \\
SAAO 0.75-m & 1764.278 & 6.53 & ERC \\
SSO 0.6-m & 1767.227 & 2.28 & RRS \\
SAAO 0.75-m & 1768.241 & 10.25 & GH \\
SAAO 0.75-m & 1769.239 & 10.27 & GH \\
LNA 0.6-m & 1769.819 & 0.24 & AB \\
SAAO 0.75-m & 1770.234 & 10.37 & GH \\
LNA 0.6-m & 1770.514 & 6.84 & AB \\
SAAO 0.75-m & 1771.237 & 10.27 & GH \\
LNA 0.6-m & 1771.498 & 3.19 & AB \\
SAAO 0.75-m & 1772.238 & 10.27 & GH \\
SAAO 0.75-m & 1773.226 & 10.49 & GH \\
SAAO 0.75-m & 1775.223 & 4.99 & GH \\
LNA 0.6-m & 1775.575 & 5.59 & AB \\
LNA 0.6-m & 1776.651 & 3.91 & AB \\
LNA 0.6-m & 1777.486 & 7.56 & AB \\
SSO 0.6-m & 1777.904 & 6.14 & RRS \\
SAAO 0.75-m & 1778.493 & 3.67 & GH \\
SAAO 1.0-m$^{\ast}$ & 1785.256 & 9.94 & GH \\
SSO 0.6-m & 1785.918 & 8.45 & RRS \\
SSO 0.6-m & 1786.991 & 0.67 & RRS \\
SSO 0.6-m & 1787.898 & 9.58 & RRS \\
SAAO 0.5-m & 1789.242 & 8.74 & GH \\
SSO 0.6-m & 1789.935 & 8.52 & RRS \\
SAAO 0.5-m & 1791.337 & 1.27 & GH \\
SAAO 0.5-m & 1792.253 & 8.42 & GH \\
SAAO 0.5-m & 1850.284 & 1.51 & DJJ \\
SAAO 0.5-m & 1852.277 & 1.78 & DJJ \\
SAAO 0.5-m & 1853.332 & 0.41 & DJJ \\
SAAO 0.5-m & 1857.270 & 2.45 & DJJ \\
SAAO 0.5-m & 1858.276 & 1.82 & DJJ \\
SAAO 0.5-m & 1859.274 & 1.92 & DJJ \\
SAAO 0.5-m & 1862.278 & 2.04 & DJJ \\
\hline
Total & & 280.35 & \\
\hline
\end{tabular}
\end{flushleft}
\end{table}
 
We used the Johnson B and V filters as well as the Cousins I$_{\rm c}$
filter (the latter only in the year 2000) with a total integration time of
$\sim$1 minute in each filter as a compromise between good time resolution
and maximum colour information. Apertures of 30 -- 45 arcseconds on the
sky were used. Sky measurements were taken at suitable intervals depending
on the brightness and proximity of the Moon.

The observing sequence was chosen to result in both good coverage for the
short-period $\delta$~Scuti pulsations and best long-term stability. We
adopted the sequence C1-V-C2-V-C1... (C1 and C2 are the comparison stars
and the V is the variable), which resulted in one variable star
measurement every $\sim$6 minutes. Supplementary observations of
UBV(RI)$_{\rm c}$ standard stars were also acquired.

Data reduction was performed in the standard way, i.e. corrections for
coincidence losses, sky background and extinction were followed by
calculating differential magnitudes between the comparison stars. The
latter were examined for variability. It seems that one of the two stars
is slightly variable on a time scale of about 1.1 or 10 days with a $V$
amplitude of 2 mmag. We investigated whether this could be due to
differential colour extinction and found no support for such an
interpretation. As our programme star exhibits long-term variability with
much higher amplitude, we cannot determine unambiguously which of the two
stars is the potential variable, but we suspect HD 209829.

In any case, we proceeded by constructing a differential target star light
curve relative to the measurements of both comparison stars. We
standardized these magnitude differences by using the transformation
equation slopes calculated from the standard star observations mentioned
above. We note that the photometric zero-points of the different
telescope/instrument combinations agreed to better than 2 mmag for each
filter or colour used during the multi-site campaign, but we experienced
some difficulties with homogenizing the data from the discovery season.
Finally, the time base of our observations was converted to Heliocentric
Julian Date (HJD). Most of the light curves obtained during the central
part of the multi-site campaign are shown in Fig.~3.

\begin{figure*}
\includegraphics[width=95mm,viewport=95 00 395 670]{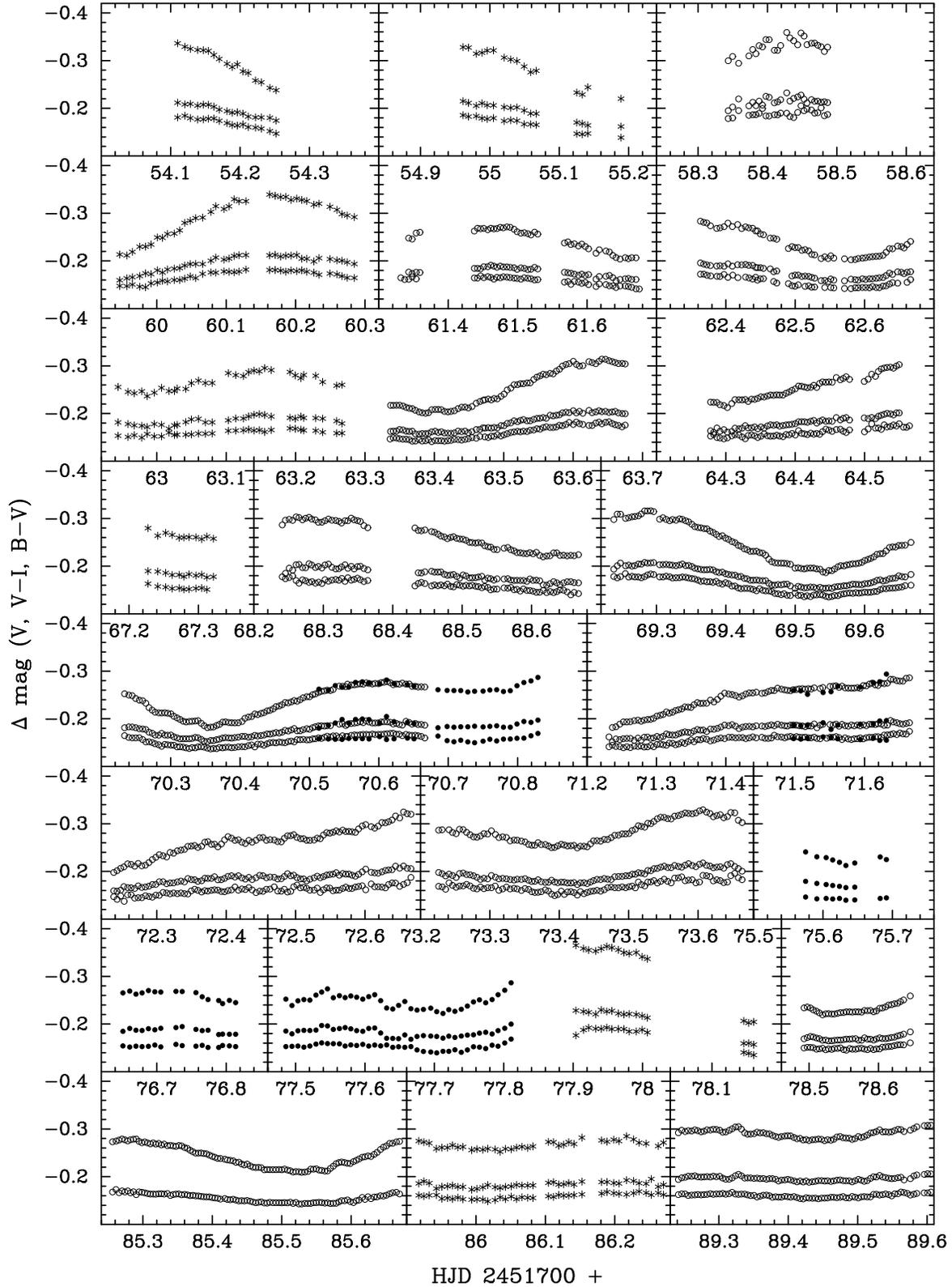}
\caption[]{Light curves of HD 209295 obtained during the multi-site
campaign in August 2000. The asterisks are measurements from SSO, the
open circles those from SAAO and the filled circles are LNA data. The
upper light curves in each panel are the $V$ data, the $(V-I_{\rm c})$
variations are shown in the middle and the lowest are the $(B-V)$ light
curves. Zero-points are relative to HD 209829; the $V$ zero-point is shifted
by $+0.11$ mag for a better display. Note the multi-periodic slow and rapid
light variability and the corresponding colour changes.}
\end{figure*}

\subsection{Infrared Photometry}

In addition to the optical data, L. A. Crause obtained infrared JHKL
measurements on the night of September 1, 2000 using the 0.75-m telescope
of the SAAO with the MkII infrared photometer (an upgraded version of the
instrument described by Glass 1973). The data were reduced to the SAAO
system by using standard stars defined by Carter (1990). As the infrared
observations were obtained simultaneously with optical photometry,
standard BVI$_{\rm c}$JHKL magnitudes could be calculated (Table 4). The
star was approaching a local minimum in its light curve when these
measurements were taken.

\begin{table}
\caption[]{Standard optical and infrared magnitudes of HD 209295 as
measured on HJD 2451792.462.}
\begin{center}
\begin{tabular}{cc}
\hline
Filter & Magnitude \\
\hline
B & 7.623 $\pm$ 0.005\\
V & 7.370 $\pm$ 0.005\\
I$_{\rm c}$ & 7.070 $\pm$ 0.005\\
J & 6.872 $\pm$ 0.009\\
H & 6.727 $\pm$ 0.007\\
K & 6.746 $\pm$ 0.008\\
L & 6.721 $\pm$ 0.081\\
\hline
\end{tabular}
\end{center}
\end{table}

These optical and infrared colours are consistent with the spectral type
of the star as inferred from standard relations (Drilling \& Landolt 2000,
Tokunaga 2000), but agree less well with the effective temperature
determined from uvby$\beta$ photometry. However, this may be due to the
high amplitude of the light variations. The total colour amplitude implies
temperature variations of $\approx500$\,K between light extrema,
sufficiently large to explain the apparent discrepancy.

\subsection{Spectroscopic observations}

Spectroscopic observations of HD 209295 were obtained with three different
telescopes. We used the 1.9-m telescope at SAAO, the 1.9-m telescope at
Mt. Stromlo (MS) in Australia, and the 1.2-m Euler telescope at the
European Southern Observatory (ESO) in Chile. A summary of these
measurements can be found in Table 5.

The bulk of the data originated from SAAO using the {\tt GIRAFFE} echelle
fibre-fed spectrograph attached to the Cassegrain focus of the 1.9-m
telescope. The {\tt GIRAFFE} spectrograph has a resolving power of about
32\,000. The $1024 \times 1024$ TEK CCD chip gives a resolution of 0.06 --
0.09~{\AA} per pixel. A Th-Ar arc lamp was used for wavelength calibration
with arc spectra taken at regular intervals to calibrate possible drifts.
The wavelength range was 4400 -- 6680~{\AA} spread over 45 orders.
Exposure times were normally 10 min for a S/N ratio of about 30 -- 60. A
total of 422 spectra of HD~209295 was obtained during three observing
runs.

The spectroscopic observations at MS were obtained with the coude
echelle spectrograph on the Mount Stromlo 1.9-m telescope. Spectra with a
resolution of $\sim0.15$ \r{A} at H$\alpha$ were recorded on a
2048$\times$4096 pixel TEK CCD. The wavelength range obtained with this
set-up was 4500 -- 6900~\r{A} spread over 42 orders. Exposure times were
between 17.5--20 min for a S/N ratio of about 25--60. A Th-Ar arc lamp was
used for wavelength calibration. Due to poor weather a total of
only 15 spectra of HD 209295 was obtained.

The spectroscopic observations in Chile were performed with the Swiss
1.2-m Ritchey-Chretien Euler telescope at ESO, La Silla. Euler is equipped
with a high-resolution echelle spectrograph, {\tt CORALIE}, and a 2k
$\times$ 2k CCD camera with 15$\mu$m pixels. The resolving power amounts
to 50\,000 and the total wavelength range is 3900\AA~-- 6800\AA~in 68
orders, without any gaps in the coverage. The {\tt CORALIE} spectra are
extracted on-line following a standard echelle reduction procedure. In the
case of our measurements, wavelength calibration utilized the most
recently obtained ThAr spectrum. Each order of the stellar spectrum was
then divided by the blaze function. For a full description of the
reduction scheme we refer to Baranne et al.\ (1996). The integration times
were adapted to the atmospheric conditions (seeing, presence of clouds)
and ranged from about 15 to 18 minutes. The typical S/N ratio was
$\approx$ 30. In total, 38 spectra were obtained during one week.

\begin{table}
\begin{center}
\caption{Log of the spectroscopic observations of HD~209295. The date with
respect to HJD~2450000, the run lengths, the mean integration times $t$,
and the number of spectra, $N$, are given.}
\begin{tabular}{lccrrr}
\hline
Site & Run start & Length & $t$ & $N$ & Observer \\
 & HJD - 2450000 & (h) & (s) & & \\
\hline
SAAO & 1501.250 & 3.02 & 600 & 16 & LAB\\
SAAO & 1502.268 & 2.39 & 600 & 13 & LAB\\ 
SAAO & 1503.262 & 2.39 & 600 & 13 & LAB\\ 
SAAO & 1504.249 & 2.99 & 600 & 16 & LAB\\ 
MS & 1505.029 & 0.33 & 1200 & 1 & DJJ\\
SAAO & 1505.255 & 2.42 & 600 & 13 & LAB\\
SAAO & 1506.250 & 0.35 & 600 & 2 & LAB\\
SAAO & 1507.297 & 1.29 & 600 & 7 & LAB\\
SAAO & 1508.249 & 2.81 & 600 & 15 & LAB\\ 
SAAO & 1509.251 & 2.54 & 600 & 12 & LAB\\
MS & 1509.960 & 0.68 & 1200 & 3 & DJJ\\
MS & 1510.924 & 2.22 & 1050 & 6 & DJJ\\
SAAO & 1511.256 & 2.42 & 600 & 13 & LAB\\ 
MS & 1511.942 & 1.62 & 1050 & 5 & DJJ\\
SAAO & 1512.254 & 2.44 & 600 & 13 & LAB\\
\hline
ESO & 1762.741 & 2.70 & 975 & 8 & TM\\
ESO & 1763.700 & 3.69 & 1100 & 10 & TM\\
SAAO & 1765.347 & 0.17 & 600 & 1 & CK\\
ESO & 1765.744 & 3.69 & 1100 & 8 & TM\\
ESO & 1766.703 & 2.92 & 1100 & 7 & TM\\
SAAO & 1768.347 & 8.08 & 1095 & 21 & CK\\
ESO & 1768.802 & 1.91 & 1100 & 5 & TM\\
SAAO & 1769.336 & 8.21 & 1016 & 24 & CK\\
SAAO & 1770.325 & 6.95 & 1000 & 21 & CK\\
SAAO & 1771.345 & 8.12 & 975 & 24 & CK\\
SAAO & 1772.345 & 7.70 & 1200 & 20 & CK\\
SAAO & 1773.354 & 7.70 & 1000 & 23 & CK\\
SAAO & 1774.406 & 1.34 & 1400 & 3 & CK\\
SAAO & 1776.328 & 0.90 & 1500 & 2 & CK\\
SAAO & 1777.388 & 1.23 & 1600 & 3 & CK\\
SAAO & 1778.348 & 5.55 & 1557 & 7 & CK\\
SAAO & 1849.267 & 2.42 & 600 & 12 & LAB \\
SAAO & 1850.271 & 2.56 & 600 & 14 & LAB\\
SAAO & 1851.274 & 2.50 & 600 & 10 & LAB\\
SAAO & 1852.262 & 2.81 & 600 & 15 & LAB\\
SAAO & 1853.258 & 2.68 & 600 & 14 & LAB\\
SAAO & 1855.354 & 0.55 & 700 & 2 & LAB\\
SAAO & 1856.258 & 2.44 & 600 & 13 & LAB\\
SAAO & 1857.265 & 2.47 & 600 & 13 & LAB \\
SAAO & 1858.264 & 2.25 & 600 & 12 & LAB\\
SAAO & 1859.263 & 2.21 & 600 & 12 & LAB\\
SAAO & 1860.265 & 2.23 & 600 & 12 & LAB\\
SAAO & 1862.262 & 2.30 & 600 & 12 & LAB\\
\hline
Total & & 128.19 & & 476 & \\
\hline
\end{tabular}
\end{center}
\end{table}

In contrast to the photometric observations, the spectroscopic
measurements were not reduced centrally. For the SAAO data, we used the
local reduction software (see {\tt http://www.saao.ac.za/facilities/}) for
corrections for for bias and flat field, order extraction and wavelength
calibration. The same reduction steps for the MS data were performed in
IRAF.

\section{Analysis}

\subsection{The photometry}

Our frequency analysis was performed with the programme {\tt PERIOD 98}
(Sperl 1998). This package applies single-frequency power spectrum
analysis and simultaneous multi-frequency sine-wave fitting, but also has
some advanced options which will be described later as necessary. We
calculated the spectral window and amplitude spectra of our data as well
as the amplitude spectra of residual light curves after the previously
identified periodicities had been removed using a multi-periodic fitting
algorithm. Similar analysis were performed for all the three filters used. 
We adopted the mean frequencies for our final solution.

Our analysis was concentrated on multi-site data from the year 2000. Due
to the different telescopes and detectors used, the 1999 data analysis is
complicated by zero-point calibration problems. Also, the data runs were
rather short in 1999 and the possibility exists that the star may have had
a different amplitude at this time. We used the 1999 data only when the
zero-point problems were relatively minor.

\subsubsection{The $\gamma$~Doradus pulsations}

In Fig.~4 we show amplitude spectra of the B filter data with consecutive 
prewhitening by the low frequencies. For the 2000 data we calculated the 
window function as the Fourier transform of a single sinusoid with frequency 
1.129 cycles/day and amplitude of 50 mmag. Due to the multi-site coverage, 
aliasing is not a problem. We detected six significant frequencies 
in the light variations. The first four frequencies were already shown in 
Table 1. 

\begin{figure}
\includegraphics[width=105mm,viewport=-10 00 345 622]{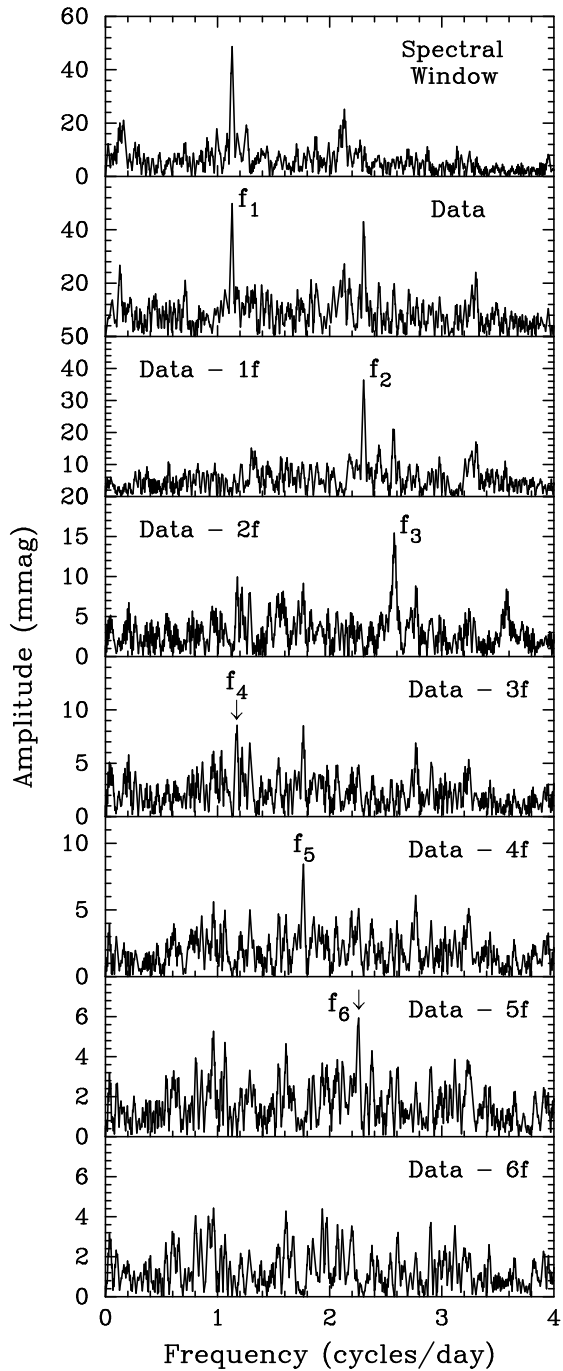}
\caption[]{Top panel: Spectral window of the B-filter time-series
photometry of HD 209295 obtained in the year 2000. The other panels show
the amplitude spectra of the data with consecutive prewhitening. Six 
significant periodicities are present.}
\end{figure}

The significance of a detection was estimated using the S/N criterion of
Breger et al. (1993). The residuals after prewhitening this frequency
solution suggest that more periodicities are present. The highest peaks in
the residual spectra are located at the same frequencies in each of the B,
V, and I$_{\rm c}$ data sets, and the ``noise level" decreases from B to
I$_{\rm c}$. This suggests that most of the ``noise" in the blue may well
be due to additional periodicities. The derived frequencies and amplitudes
are listed in Table 6.

\subsubsection{The $\delta$~Scuti pulsations}

The zero point uncertainties in the data set from 1999 have little effect
in the frequency range in which the $\delta$~Scuti pulsations are
present. Consequently, we can incorporate these observations into the
frequency analysis. We therefore prewhitened the data by the six-frequency 
solution discussed above from the 2000 data. For the 1999 data we
prewhitened by the first five of these frequencies which were found in
common with the analysis of the 2000 data. This prewhitening is important
because low-frequency variations can artificially increase the noise level in 
the high-frequency domain through spectral leakage. The amplitude spectrum of
the combined B filter data and subsequent prewhitening are shown in
Fig.~5.

\begin{figure}
\includegraphics[width=100mm,viewport=-10 00 335 260]{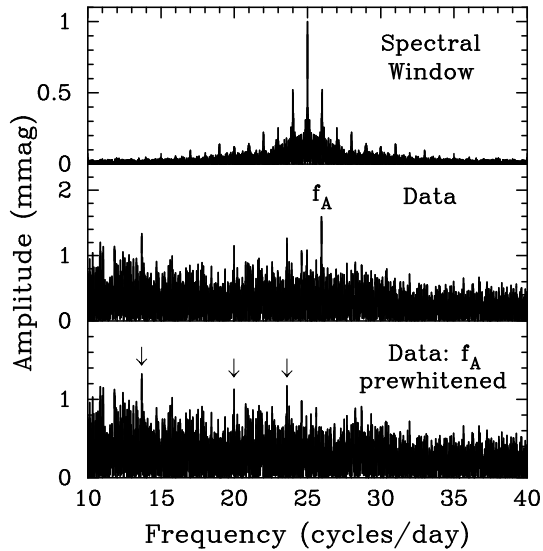}
\caption[]{The spectral window of the combined residual B filter data and
amplitude spectra in the frequency domain of the $\delta$~Scuti
pulsations. Only one frequency is convincingly detected, but the presence
of several low-amplitude modes in a range of 12 -- 25 cycles/day is
suspected; some possible further frequencies are indicated.}
\end{figure}

Although several peaks in the amplitude spectra in Fig.~5 are prominent,
only one signal is a significant detection. This is surprising, as typical
multi-periodic beating is seen in the light curves of Fig.~3, e.g.
HJD~2451770.35 vs. HJD 2451770.55. We therefore checked whether these
signals could be non-coherent by analysing several subsets of data. No
evidence for short-lived $\delta$~Scuti type variations or variations with
changing amplitude was found. In fact, several of the more conspicuous
signals were present in all data subsets (e.g. in different subsets of the
multi-site data or in individual filters). Thus we suspect that several
such periodicities may be real. A multi-frequency solution with $f_A$ and
the three next strongest variations in the $\delta$~Scuti regime fits the
short-term variations reasonably. The corresponding peaks are marked with
arrows in the lowest panel of Fig.~5. The frequency range of possibly
excited $\delta$~Scuti pulsations can be constrained to 10 -- 32
cycles/day. We list the results of our frequency analysis of the
photometric data in Table~6.

\begin{table*}
\caption[]{Results of a multi-frequency solution for HD 209295 derived
from the photometric data. Frequency error estimates range from $\pm$
0.0001 cycles/day to $\pm$ 0.002 cycles/day, for the strongest and weakest
$\gamma$~Dor mode respectively, and $\pm$ 0.005 cycles/day for the
$\delta$~Scuti mode. Signal-to-noise ratios following Breger et 
al. (1993), calculated from the B filter data from the year 2000 are also
given. A $S/N>4$ is taken to be a significant detection of a signal.}
\begin{center}
\begin{tabular}{cccccccc}
\hline
 & & & \multicolumn{3}{c}{{\bf Measurements from the year 2000}} &
\multicolumn{2}{c}{{\bf Measurements from the year 1999}} \\
ID & Frequency & $S/N$ & $B$ Amplitude & $V$ Amplitude & $I$ Amplitude &
$B$ Amplitude & $V$ Amplitude \\
 & (d$^{-1}$) & & (mmag) & (mmag) & (mmag) & (mmag) & (mmag) \\
 & & & $\pm$ 0.4 & $\pm$ 0.3 & $\pm$ 0.2 & $\pm$ 0.6 & $\pm$ 0.5\\
\hline
\multicolumn{8}{c}{{\bf $\gamma$~Doradus frequencies}} \\
\hline
$f_1$ & 1.1296 & 35.9 & 49.6 & 38.5 & 23.0 & 35.9 & 26.9 \\
$f_2$ & 2.3024 & 28.3 & 39.1 & 28.3 & 15.5 & 24.5 & 17.3\\
$f_3$ & 2.5758 & 11.6 & 16.0 & 11.4 & 6.2 & 21.7 & 15.8\\
$f_4$ & 1.1739 & 9.0 & 12.4 & 7.9 & 4.6 & 5.4 & 4.9 \\
$f_5$ & 1.7671 & 7.0 & 9.7 & 7.4 & 4.1 & 11.0 & 8.5 \\
$f_6$ & 2.2572 & 5.4 & 7.4 & 5.3 & 3.1 & n/a & n/a \\
\hline
\multicolumn{8}{c}{{\bf $\delta$~Scuti frequency}} \\
\hline
$f_A$ & 25.9577 & 5.3 & 1.8 & 1.4 & 0.8 & 1.5 & 1.3 \\
\hline
\end{tabular}
\end{center}
\end{table*}

We need to make two remarks here. Firstly, the results for the data from
1999 are only listed for completeness in Table 6. We suspect that the
amplitudes of the $\gamma$~Dor pulsations in this data set are
artificially decreased not only because of zero-point problems, but also
due to poor data sampling. Secondly, we stress that these are only
preliminary results. We will revisit the frequency analysis in Section
3.4.

\subsection{Spectroscopic analyses}

The first step in the analysis is to rectify the spectra, i.e. to place
the continuum. This was done by using an unbroadened synthetic spectrum
with T$_{\rm eff} = 7500$~K, $\log g = 4.00$ as a template, using the {\tt
SPECTRUM} code (Gray \& Corbally 1994). A running median of each echelle
order was divided by the corresponding section of the synthetic spectrum
and taken to represent the response function of the instrument. A
polynomial of degree 5 was fitted to the response function and used to
correct the observed spectrum. The result is the rectified spectrum of the
star used for cross-correlation.

\subsubsection{Radial velocities}

For each order, the observed rectified spectrum was correlated with the
corresponding section of the synthetic spectrum after removing the unit
continuum. The resulting correlation function is, in effect, the mean line
profile with the continuum removed. The ``radial velocity'' for each order
is obtained by fitting a quadratic to the correlation function and finding
the position of the maximum. The mean radial velocity from all the orders
is also obtained. A standard error of typically 1 -- 3 km~s$^{-1}$ was
found for the SAAO spectra.

A few of the spectroscopic observations listed in Table 5 turned out to be
of a quality too poor for further use. Four spectra were discarded, and
472 were retained. Some slight deviations ($<0.5$ km/s) in the seasonal
radial velocity zero-points in the SAAO spectra were rectified by means of
telluric lines.

An immediate finding from our time-resolved spectroscopy is that the
radial velocities of HD 209295 are strongly variable. The total radial
velocity amplitude is in excess of 100 km/s, which cannot be due to the
pulsations of the star. HD 209295 must be a member of a binary system.

We therefore attempted to determine the orbital period. Visual inspection
of the radial velocity curve suggests $P_{\rm orb} \sim$ 3 days, and
Fourier analysis (upper panel of Fig. 6) indeed implies an orbital
frequency near 0.32 cycles/day, but our data set is affected by aliasing.
Prewhitening trial frequencies suggested that the shape of the orbital
radial velocity curve is not sinusoidal.

This is a situation in which use of the residualgram method (Martinez \&
Koen 1994) is indicated, which is based on a least-squares fit of a sine
wave with $M$ harmonics. The residual sum of squares at each trial
frequency is evaluated; $M$ can be chosen freely. In that way, alias
ambiguities can be circumvented by taking advantage of the information in
the Fourier harmonics. Preliminary trials with Fourier analysis suggested
that $M=3$ is a good choice for a residualgram as shown in the lower panel
of Fig. 6.

\begin{figure}
\includegraphics[width=112mm,viewport=00 00 368 344]{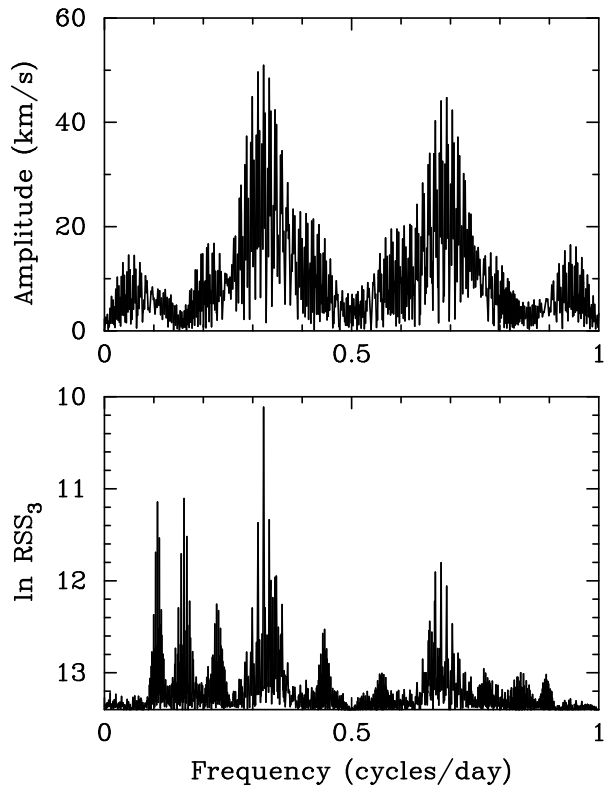}
\caption[]{Upper panel: Fourier amplitude spectrum of the radial velocities
of HD 209295. An orbital period near 3 days is implied, but
severe aliasing is present. Lower panel: residual sum of squares spectrum
of a 3-harmonic fit (RSS$_3$) to the radial velocities. Although some
power leaks into sub-harmonics, the correct orbital frequency is now
unambiguously detected.}
\end{figure}

The residualgram method eliminated the aliasing problem for the
determination of the orbital frequency. We use its result for a refinement
of the orbital frequency by using {\tt Period 98} (Sperl 1998). This
leads to an improved orbital frequency of $f_{\rm orb} = 0.32198$
cycles/day ($P_{\rm orb}$ = 3.10575 d).

To determine the orbital solution from our radial velocities, we firstly
used the SAAO data only, which are the most extensive (almost 90\% of all
spectra) and are homogeneous. We weighted the individual measurements
based on their standard errors. Measurements with standard errors smaller
than 2 km/s were given a weight of 1; the weights decreased down to 0.2
for a few measurements with standard errors between 8--9 km/s. We
determined an initial orbital solution from these data with an updated
version (Strassmeier, private communication) of the
differential-correction method (Barker, Evans \& Laing 1967).

Using the initial parameters from these methods, we examined the relative
zero-points of the three blocks of SAAO measurements; no third body in the
system was detected. We then fitted all the campaign data with these
orbital parameters and examined the zero-points of the measurements from
the other sites. The ESO radial velocities needed to be shifted by $+7$
km/s. We then calculated our final solution from all the data, with the
mean radial velocity fixed. This orbital solution is shown in Table 7 and
Fig. 7.

\begin{table}
\caption{The orbital solution from radial velocities of HD 209295. The
symbols have their usual meanings; $\omega$ is the argument of
periastron.}
\begin{center}
\begin{tabular}{ccc}
\hline
Parameter & Unit & Value \\
\hline
$P_{\rm orb}$ & (days) & 3.10575 $\pm$ 0.00010 \\
$\gamma$ & (km/s) & -23.7 $\pm$ 0.4 \\
$K$ & (km/s) & 54.2 $\pm$ 0.7\\
$e$ & & 0.352 $\pm$ 0.011 \\
$\omega$ & (degrees) & 31.1 $\pm$ 2.0 \\
$T_0$ & HJD & 2451771.864 $\pm$ 0.014 \\
$a_1 \sin i$ & $R_{\sun}$ & 3.11 $\pm$ 0.04 \\
$f(m)$ & $M_{\sun}$ & 0.042 $\pm$ 0.002\\
\hline
\end{tabular}
\end{center}
\end{table}

\begin{figure}
\includegraphics[width=105mm,viewport=-08 00 340 176]{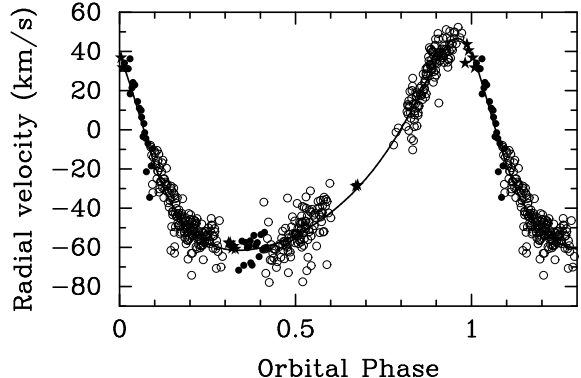}
\caption[]{Radial velocities of HD 209295 phased with the orbital solution
(solid line). Open circles are SAAO radial velocities, filled circles are
ESO radial velocities and star symbols denote the MS measurements. We note
that although the number of spectra from ESO and MS is small, they cover
important phases of the orbit and are therefore quite valuable. The formal
radial velocity errors quoted in Sect. 3.2.1 underestimate the true
errors.}
\end{figure}

We note that an extrapolation of our orbital solution to the measurements
of Grenier et al. (1999) quoted in the introduction results in a good fit.
The two measurements by these authors were taken at similar orbital
phases, which explains the comparably small radial velocity difference.

Having solved for the orbital parameters, we can now discuss two issues:
the nature of the companion and the possibility of mass transfer in the
system. In the upper panel of Fig. 8., we show the dependence of the
secondary mass on orbital inclination for various values of the primary
mass. From this, we see that $M_2>0.62M_{\sun}$. The lower panel of Fig. 8
shows that HD 209295 may come close to filling its Roche Lobe (calculated
with the approximation by Eggleton 1983) at periastron if the companion
has a small mass. However, we will show in Sect. 4.1 that the companion
is probably too massive for the primary to fill its Roche Lobe.

Assuming that the companion of HD 209295 is a main-sequence star, its
absolute magnitude $M_{\rm v} > 4.4$; otherwise its spectral lines would
be detected. A spectral type of G0 and later is therefore suggested with a
mass $<1.05M_{\sun}$ (Drilling \& Landolt 2000). Our photometry also shows
no eclipses (within a conservative limit of 4 mmag), allowing us to infer
the constraint $i<80\degr$. A main-sequence companion of
$0.62M_{\sun}<M_2<1.05M_{\sun}$ should be detected by means of its
infrared excess. We will examine this in Sect. 3.5.

\begin{figure}
\includegraphics[width=105mm,viewport=-08 00 367 369]{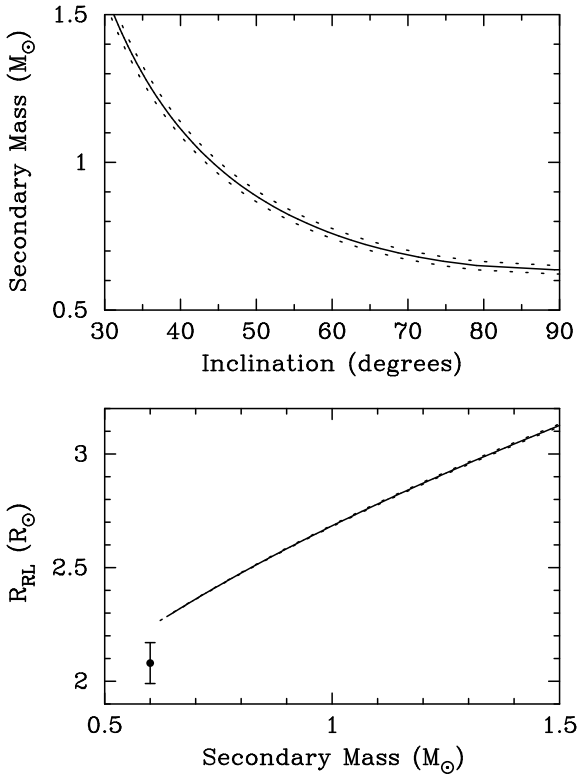}
\caption[]{Upper panel: the mass of the companion of HD 209295 versus
orbital inclination. The solid line is for a primary mass of 1.84
$M_{\sun}$ as inferred in Sect. 1.1, the dotted lines represent the
1$\sigma$ limits of this determination. Lower panel: the Roche Lobe radius
of the HD 209295 primary at periastron depending on secondary mass, again
for $M=1.84 \pm 0.07 M_{\sun}$. The filled circle with the error bar shows
our radius determination for HD 209295 from Sect. 1.1.}
\end{figure}

After removal of the orbital variation, the residual radial velocities can
be searched for pulsational signals. The amplitude spectrum of these
residuals can be found in Fig. 9. Although there are signals in the same
frequency range as in the photometry, none gives a reliable detection
according to the signal-to-noise criterion of Breger et al. (1993).
Consequently, we need to continue the examination of the spectroscopic
evidence of the pulsations of HD 209295 by looking at the line-profile
variations.

\begin{figure}
\includegraphics[width=105mm,viewport=-08 00 352 170]{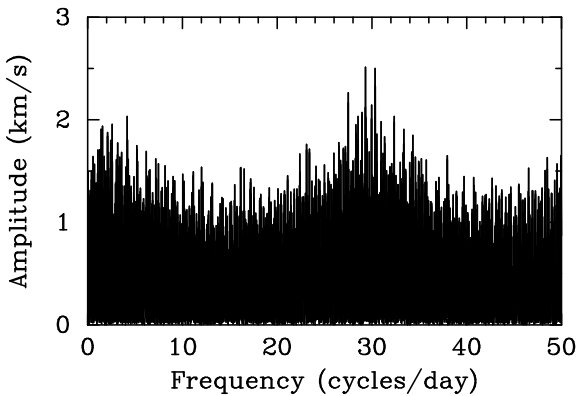}
\caption[]{The amplitude spectrum of the residual radial velocities of HD
209295 after removing the orbital solution.}
\end{figure}

\subsubsection{Projected rotational velocity}

The correlation profiles for each season can be co-added (after correcting
with our orbital solution) to form what is essentially a mean line
profile. The projected rotational velocity, $v \sin i$, can be determined
by fitting a model profile of a rotating star. The model is spherical with
a linear limb darkening coefficient $u = 0.3$ and the visible hemisphere
is divided into a large number of surface elements. The intrinsic line
profile for each element summed over the hemisphere, weighted according to
limb darkening, is computed. The projected rotational velocity and mean
radial velocity are adjusted until a best fit is obtained to the observed
profile. We obtain $v \sin i = 108 \pm 10$~km~s$^{-1}$ for the SAAO
measurements from 1999 and the second set from 2000, and $v \sin i = 98
\pm 10$~km~s$^{-1}$ for the first data set of the year 2000. Because of
the uncertainty in the wings of the correlation profile, the fit was
performed using the part of the profile within 90~km~s$^{-1}$ of the line
centre.

\subsubsection{The high-order profile variations}

In a recent paper on the $\delta$~Sct star 38~$o^1$~Eri, Balona (2000) was
able to show that pulsation modes of high spherical degree $\ell$ are easy
to resolve in stars with moderate to high rotational velocities. Bearing
in mind the result from the previous paragraph, HD~209295 falls within
this category. We will only use the SAAO spectra for the following
analysis, as these are the most homogeneous and best sampled.

Line profile variations due to modes of high degree are seen as moving
sub-features in the correlation profiles of the spectra used to determine the 
radial velocities. As a particular sub-feature crosses the centre of the
profile, it determines the radial velocity. A short while later another
traveling sub-feature will appear and the radial velocity will suddenly
change. These discontinuous jumps are partly responsible for the rather
high errors of the radial velocities.

From the spacing of the traveling sub-features, we estimate that the
spherical degree is $\ell \approx 5$. To enhance the profile variations,
we removed the orbital motion and constructed the mean correlation profile
from the SAAO data. Each correlation profile was divided by the mean
profile to construct ``difference'' profiles at the given times. The
moving features are most clearly seen when these difference profiles are
arranged in a time sequence and converted to a grey-scale image. An example
is shown in Fig.~10.

\begin{figure}
\includegraphics[width=115mm,viewport=15 00 477 342]{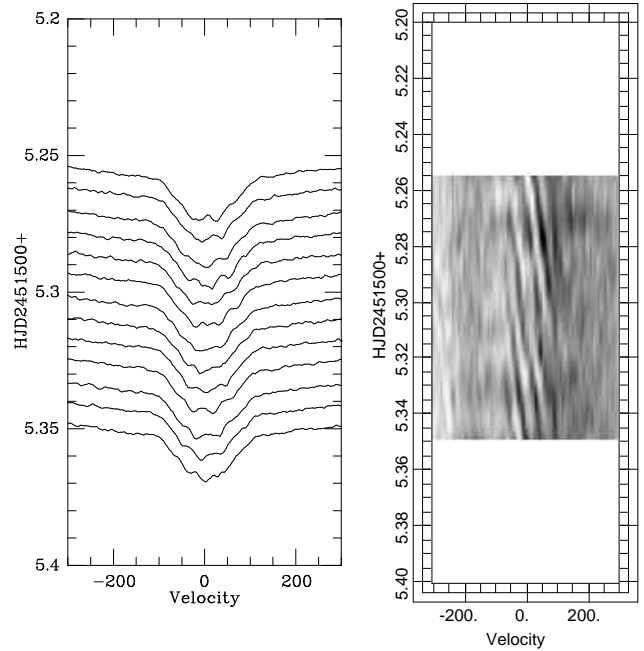}
\caption[]{Left panel - correlation function; right panel - grey-scale
difference after removing the correlation function. The data are from the
1999 season. The abscissa is in km~s$^{-1}$ and the ordinate is the HJD
measured from HJD~2451500.000.}
\end{figure}

While it is clear from Fig.~10 that periodic high-order line profile
variations are present, it is not possible to determine the frequencies
from the radial velocities alone, as shown before. Clearly, the radial
velocity is a poor indicator for modes of high degree.

A common method to determine periodicities in line profile variations is
to calculate the periodogram along wavelength bins throughout the line
profile, or across the correlation profile. Most of the signal in this
analysis is located at frequencies very close to the orbital period and
its aliases, suggesting that the removal of the orbital variation did not
work satisfactorily. However, we found the signatures of two previously
detected modes in the line profiles, although they cannot be detected
without prior knowledge. These are the photometric $\gamma$ Doradus modes
$f_1$ and $f_2$.

This finding gives us the possibility to attempt a mode identification
with the method proposed by Telting \& Schrijvers (1997). This technique
utilises the phase change of a mode as it progresses through the line
profile, and provides some simple mode identification diagnostics
depending on the total phase change of the mode frequency and its first
harmonic (e.g. $\Delta \phi_0 \approx \pi\ell$). We calculated these phase
changes for the two modes which we think are present and we show their
behaviour through the correlation profile in Fig.~11. The mode
identification implied by this analysis is summarized in Table~8.

\begin{figure}
\includegraphics[width=110mm,viewport=0 00 345 266]{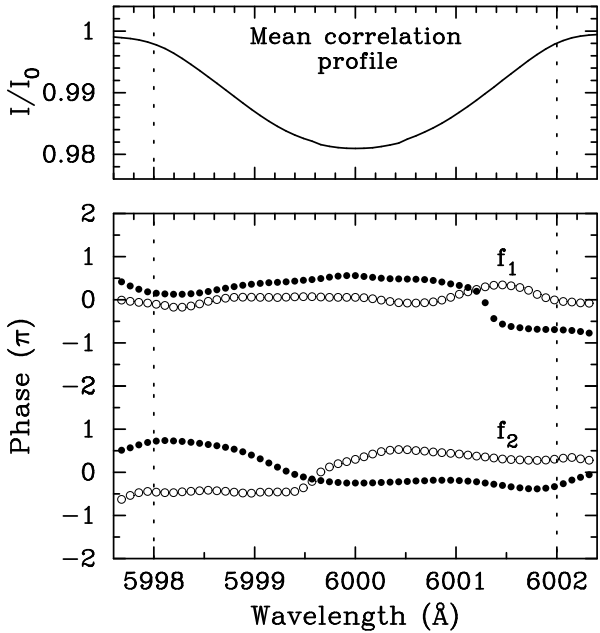}
\caption[]{Phase diagram for two modes of HD 209295 (lower panel) through
the correlation profile (upper panel), which is arbitrarily centred at
6000\AA. The full circles denote the phases of the mode frequency, and the
open circles that of its first harmonic. The zero-points in the phases are
arbitrary. The dotted vertical lines indicate $v \sin i$ as determined in
Sect. 3.2.2 translated into units of wavelength. This is also the position
in the correlation profile where we determine the total phase change.}
\end{figure}

\begin{table}
\caption{The total phase variation of the mode frequencies $\Delta \phi$,
their first harmonics $\Delta \phi_1$, and the implied mode
identifications following Telting \& Schrijvers (1997).}
\begin{center}
\begin{tabular}{ccccc}
\hline
ID & $\Delta \phi_0$ & $\Delta \phi_1$ & $\ell$ & $|m|$\\
 & ($\pi$ rad) & ($\pi$ rad) & & \\
\hline
$f_1$ & 0.85 & $-0.05$ & 1 & 1 \\
$f_2$ & 1.06 & $-0.77$ & 1 & 1 \\
\hline
\end{tabular}
\end{center}
\end{table}

We note that Telting \& Schrijvers (1997) have given error estimates for
the mode identifications with this method with $\pm 1$ for $\ell$ and $\pm
2$ for $m$. The identifications are plausible: $f_1$ and $f_2$ have
high photometric amplitude.

\subsection{Combining the spectroscopic and photometric information}

Comparing the frequencies of the $\gamma$~Doradus pulsations in Table 6 to
the orbital frequency we see that the photometric mode $f_3$ corresponds 
exactly to 8$f_{\rm orb}$, and $f_6$ is, within the errors, consistent with 
7$f_{\rm orb}$. As noted in Sect. 1.1, the relation $f_4=f_2-f_1$ is 
also present within the mode frequencies.

Consequently, we revisited the frequency analysis of the photometric data.
We again restricted ourselves to the data from 2000. This time, however,
we used {\tt Period 98}'s capability to fix signal frequencies to certain
values, in our case integer multiples of the orbital frequency, and to
perform nonlinear least squares fits with frequencies thus fixed.

We started by using the six $\gamma$~Doradus frequencies in Table 6, and
proceeded by the usual prewhitening procedure. The corresponding amplitude
spectra can be found in Fig.~12; they are generated by combining the B
filter residuals with the V filter residuals multiplied by a factor of
1.29.

\begin{figure}
\includegraphics[width=110mm,viewport=-05 00 337 570]{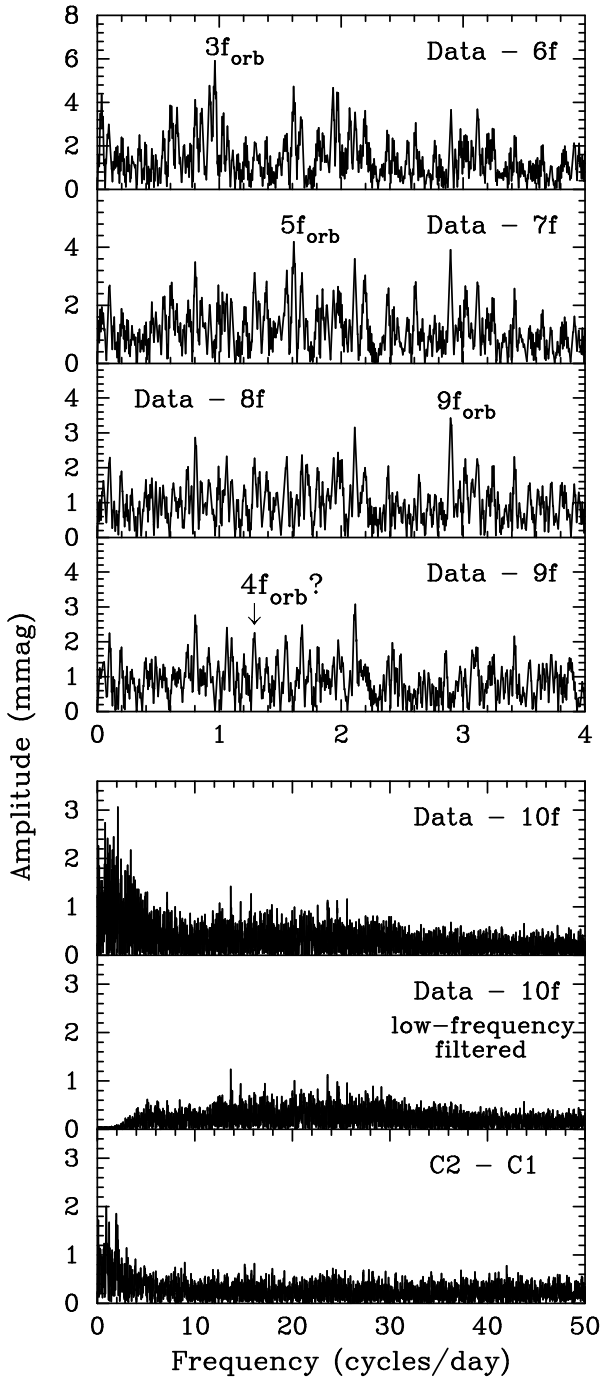}
\caption[]{Top panel: the amplitude spectrum of the combined B and scaled
V-filter residuals of HD 209295 after subtraction of the frequency
solution of Table 6. Another harmonic of the orbital frequency is
detected. In the following three panels we show the effects of further
prewhitening. All new detections are again harmonics of $f_{\rm orb}$.  
Third panel from bottom: the residual amplitude spectrum in a wider
frequency range. Second panel from bottom: the residual amplitude spectrum
after filtering the residual low-frequency variations. Bottom panel:
amplitude spectrum of the differential magnitudes of the comparison
stars.}
\end{figure}

The highest-amplitude peak in the upper panel of Fig.~12 does, indeed,
correspond to 3$f_{\rm orb}$. In addition 5$f_{\rm orb}$ (second panel of
Fig.~12) and 9$f_{\rm orb}$ (third panel of Fig.~12) are also found.  
This latter frequency corresponds to exactly $f_1+f_5$. We therefore fixed
this frequency combination to its exact value before proceeding. After
prewhitening these frequencies, some peaks, including one at 4$f_{\rm
orb}$, are still visible in each filter, but not at a level considered
significant.

The third panel from the bottom of Fig.~12, shows the residual amplitude
spectra in a wider frequency range after prewhitening by the
$\delta$~Scuti mode, $f_A$. Again, there is evidence of additional
$\gamma$~Doradus and weak $\delta$~Scuti periodicities. The referee
suggested to filter out the slow variability to examine the claimed
$\delta$~Scuti variability further. This is done in the second panel from
the bottom of Fig.~12; albeit somewhat affected by the filtering, the
additional $\delta$~Scuti variations remain with similar amplitudes. They
are therefore not artifacts from spectral leakage of the low frequencies.
In the lowest panel of Fig.~12 we show the amplitude spectrum of the
comparison star data. Although the number of observations of each
comparison star is half that of the program star, the noise level in this
periodogram is lower still, again suggestive of further periodicities
below the detection level.

In Table 9 we show the final multi-frequency solution for the 2000 data set.
The corresponding results for 1999 are of little value, as the data sampling 
is too poor to attempt such a fit. Error estimates are formal values following
Montgomery \& O'Donoghue (1999). While they should be quite realistic for the
$\delta$~Sct modes, they probably underestimate the errors for the $\gamma$ 
Dor modes by a factor of 3--4.

\begin{table*}
\caption[]{The final multi-frequency solution for HD 209295 from all the
year 2000 photometry. Pulsational phases for mean light level 
are given with respect to a time of periastron passage, HJD 2451771.864.
Amplitude signal-to-noise ratios are calculated from Fig. 12.}
\begin{center}
\begin{tabular}{cccccccccc}
\hline
ID & Combination & Frequency & $S/N$ & $B$ Ampl. & $B$ phase & $V$
Ampl. & $V$ phase & $I$ Ampl. & $I$ phase \\
 & & (d$^{-1}$) & & (mmag) & ($\degr$) & (mmag) & ($\degr$) & (mmag) &
($\degr$)\\
 & & & & $\pm$ 0.3 & & $\pm$ 0.2 & & $\pm$ 0.2 & \\
\hline
\multicolumn{10}{c}{{\bf $\gamma$~Doradus frequencies}} \\
\hline
$f_1$ & & 1.12934 $\pm$ 0.00005 & 63.7 & 50.3 & -39.4 $\pm$ 0.3 & 38.9 &
-40.3 $\pm$ 0.3 & 23.5 & -42.6 $\pm$ 0.4\\
$f_2$ & & 2.30217 $\pm$ 0.00006 & 49.9 & 39.4 & -115.4 $\pm$ 0.4 & 28.7 &
-116.4 $\pm$ 0.4 & 15.4 & -117.0 $\pm$ 0.6 \\
$f_3$ & 8$f_{\rm orb}$ & 2.57593 $\pm$ 0.00011 & 23.2 & 18.3 & 66.5 $\pm$
0.8 & 13.2 & 67.2 $\pm$ 0.9 & 7.3 & 64.3 $\pm$ 1.3 \\
$f_4$ & $f_2-f_1$ & 1.17283 $\pm$ 0.00004 & 14.4 & 11.4 & -80.4 $\pm$ 1.3
& 7.6 & -73.3 $\pm$ 1.5 & 4.0 & -68.5 $\pm$ 2.4 \\
$f_5$ & 9$f_{\rm orb}-f_1$ & 1.76859 $\pm$ 0.00005 & 13.6 & 10.8 & 96.5
$\pm$ 1.3 & 8.2 & 95.4 $\pm$ 1.4 & 4.8 & 86.2 $\pm$ 2.0 \\
$f_6$ & 7$f_{\rm orb}$ & 2.25394 $\pm$ 0.00011 & 10.6 & 8.4 & 2.0 $\pm$
1.7 & 6.6 & -1.0 $\pm$ 1.8 & 3.5 & 2.0 $\pm$ 2.8 \\
$f_7$ & 3$f_{\rm orb}$ & 0.96597 $\pm$ 0.00011 & 8.9 & 7.0 & -39.2 $\pm$
2.0 & 6.2 & -35.9 $\pm$ 1.9 & 4.9 & -33.3 $\pm$ 1.9 \\
$f_8$ & 5$f_{\rm orb}$ & 1.60996 $\pm$ 0.00011 & 5.8 & 4.6 & -162.1 $\pm$
3.1 & 3.9 & -158.6 $\pm$ 3.0 & 1.9 & -173.9 $\pm$ 5.2 \\
$f_9$ & 9$f_{\rm orb}$ & 2.89792 $\pm$ 0.00011 & 5.7 & 4.5 & 47.0 $\pm$
3.2 & 3.5 & 48.9 $\pm$ 3.3 & 2.2 & 53.5 $\pm$ 4.3\\
\hline
\multicolumn{10}{c}{{\bf $\delta$~Scuti frequency}} \\
\hline
$f_A$ & & 25.9577 $\pm$ 0.0015 & 5.2 & 1.8 & 158.9 $\pm$ 8.3 & 1.4 & 162.8
$\pm$ 8.3 & 0.8 & 179.1 $\pm$ 12.2 \\
\hline
 & $f_{\rm orb}$ & 0.32199 $\pm$ 0.00011 \\
\hline
\end{tabular}
\end{center}
\end{table*}

The rms residual errors per single data point for this solution are 7.0
mmag in the B filter, 5.7 mmag in the V filter and 4.5 mmag in the I
filter. These values again suggest undetected periodicities, since the rms 
scatter of a single comparison star measurement is 5.1 mmag in B, 4.0 mmag in 
V and 4.2 mmag in I.

Another estimate of the orbital period can be made from the frequency
analysis. We find $P_{\rm orb} = 3.1057 \pm 0.0010$, consistent
with the determination in Sect. 3.2.1. We have examined the residual light
curve for possible ellipsoidal variations, but found none exceeding
2 mmag. Finally, one might suspect that $f_6=2f_1$, but a fit assuming
this relation gives a significantly poorer solution than the one
listed in Table~9.

\subsection{Attempts at mode identification from colour photometry}

The main reason for observing HD 209295 in more than one filter was to
attempt a mode identification. This method relies on comparing theoretical
amplitude ratios and phase differences of the variations in different
wavebands with the observed values. The high amplitude of the light
variations of HD 209295 make the star very well suited for this method.
Because of the very low amplitude of the $\delta$~Scuti pulsations, only
the $\gamma$~Doradus modes are considered. However, our calculations
indicate that the amplitude ratios and phase differences for the
$\delta$~Scuti mode are consistent with pulsation.

We first attempted to apply the method developed by Koen et al. (1999) to
the $\gamma$~Doradus modes. This technique uses the observed amplitude
ratios and phase differences in all filters simultaneously. Unfortunately,
no meaningful results were obtained. We then constructed two-colour
diagnostic diagrams showing amplitude ratio as a function of phase
difference (Watson 1988). In such diagrams, the theoretically determined
areas of interest are specified and compared with the observations. We
used the Warsaw-New Jersey stellar evolution and W.A. Dziembowski's NADROT
pulsation code for models of 1.8 and 1.9 $M_{\sun}$ with effective
temperatures between 7250 and 7800 K. These models span the possible range
of parameters for HD 209295. We then computed theoretical amplitude ratios
and phase shifts for the eigenmodes of those models following Balona \&
Evers (1999) and examined their location in the corresponding diagrams. We
noticed that (except for a few combinations of $Q$ and $\ell$) the results
clustered in well-defined regions depending on $\ell$. Hence, we defined
those regions as our regions of interest and compared them to the
observations. An example is shown in Fig. 13. 

\begin{figure}
\includegraphics[width=110mm,viewport=05 05 337 515]{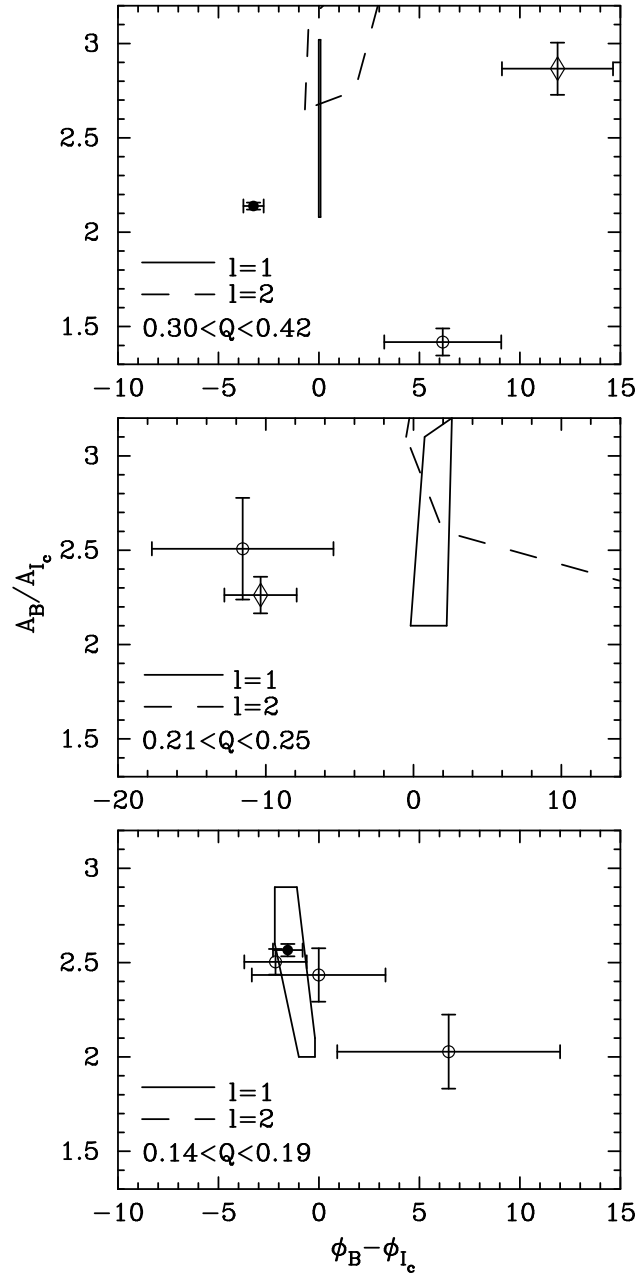}
\caption[]{Comparison of theoretical and observed B/I$_{\rm c}$ colour
amplitude ratios and phase differences with observations. Upper panel:
variations with frequencies less than 1.2 cycles/day. Middle panel:
variations with frequencies between 1.6 to 1.8 cycles/day. Lower panel:
variations with frequencies greater than 2.2 cycles/day. The areas
outlined by the full lines are theoretical predictions for $\ell=1$, and
areas represented by dashed lines correspond to $\ell=2$. Symbols with
error bars are the observed values. Filled circles correspond to
independent modes, open circles to harmonics of the orbital period and
diamonds to combination frequencies. Note the different abscissa scale in
the middle panel. Regions of interest were also calculated for $\ell=3$,
but they are off scale towards higher amplitude ratios. The $\ell=2$ area
for $0.14<Q<0.19$ is off scale towards larger positive phase shifts.}
\end{figure}

Before evaluating the potential for mode identifications from Fig.~13, two
comments need to be made. Firstly, we only show the results for the B and
I$_{\rm c}$ filter pair, as the diagrams are similar in appearance for the
other two possible filter combinations, but the observational errors are
larger.

Secondly, the locations of the areas of interest depend quite strongly on
the pulsation constant $Q$. This is a well-known result (e.g. discussed by
Garrido 2000), and is the reason why we have separated the results into
three subsets which depend on $Q$. We chose a suitable subdivision based
on the observed $\gamma$~Dor frequencies.

The implications from Fig.~13 can be summarized as follows. For modes
with $Q>0.2$~d, the mode identification method gives meaningless results.
For the modes with $Q<0.2$~d there is some agreement, but this might be
coincidence. In any case, the amplitude ratios of the photometric modes
are consistent with $\ell=1$ or 2, but not with $\ell=3$.

\subsection{The infrared colours and the companion of HD 209295}

In Section 3.2.1 we have shown that the binary companion of HD 209295 must
have a mass of at least 0.6 $M_{\sun}$. If it were a main-sequence star,
this would correspond to a spectral type of K5. The most luminous possible
main-sequence companion to HD 209295 is a G0 star. A star more luminous
than this would be detected in the spectrum. We decided to search for a
companion in the infrared. For this purpose we show the system's optical
and infrared colours in Table 4. We note that interstellar reddening is
insignificant because of the proximity of the star ($d=122$ pc) and its
high galactic latitude ($b=-44\degr$); there is also no evidence for
reddening in the observed Str\"omgren colours (Table 2).

We adopted standard relations for absolute magnitude as well as optical
and infrared colours of main-sequence stars from the tables of Tokunaga
(2000) and Drilling \& Landolt (2000). Given the observed $V$ magnitude
and $(V-I_{\rm c})$ colour of the system, we added the fluxes of
hypothetical G0, K0 and K5 main-sequence companions. For the primary's
flux we used a suitable interpolation between A7 and F0 to reproduce the
observed $V$ and $(V-I_{\rm c})$. The infrared colours of those models
were then compared to observations. The result, displayed in Fig.~14,
shows the relative infrared excess of the observations and of some of the
binary models compared to the expected infrared magnitudes.

\begin{figure}
\includegraphics[width=92mm,viewport=-6 10 312 200]{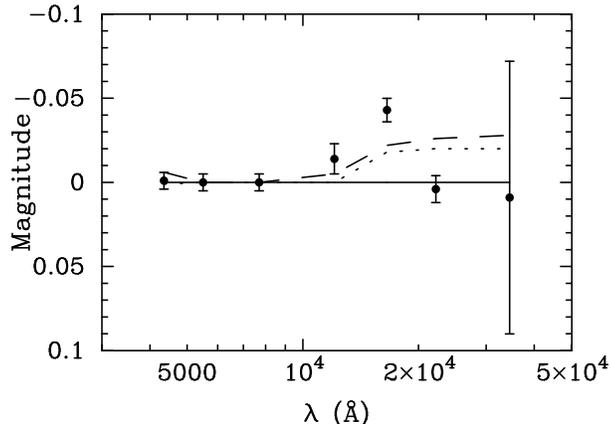}
\caption[]{A search for infrared excess in the energy distribution of HD
209295. The solid horizontal line represents an interpolation of standard
relations for A-type stars to the observed $(V-I_{\rm c})$ colour of HD
209295. The dotted line is the same with a G0 main-sequence companion
added, and the dashed line shows the expected colours if a K5
main-sequence companion were present. A K0 companion would be practically
indistinguishable from a G0 star in this diagram. The filled circles with
the error bars are the actual observations.}
\end{figure}

Whereas the infrared $J$ and $H$ magnitudes seem to indicate the presence
of a companion, the $K$ magnitude argues against this idea; the $L$
magnitude is too uncertain to be useful. The infrared excess we calculated
also depends on the uncertainties of the standard relations and the
accuracy of our measurements. We conclude that we cannot pinpoint the
nature of the companion of HD 209295 from the infrared data alone.

\section{Further discussion of the observations}

\subsection{The secondary component and evolutionary history of the HD
209295 system}

We were unable to detect the secondary component of the HD 209295 system
in the infrared. However, we can still obtain further constraints on its
nature. The first constraint is the absence of ellipsoidal variability.

Morris (1985) derived theoretical expressions for system parameters
of ellipsoidal variables from the observed light curves. Using his Eq. 6,
a limb darkening coefficient of 0.59 from Claret (2000), a gravity
darkening exponent of 0.84 (Claret 1999), and a conservative upper limit
of 5 mmag on the peak-to-peak amplitude of possible ellipsoidal variations
of HD 209295, we can determine the minimum orbital inclination of the system
as a function of secondary mass. The result is shown in Fig.~15.

\begin{figure}
\includegraphics[width=97mm,viewport=00 00 312 177]{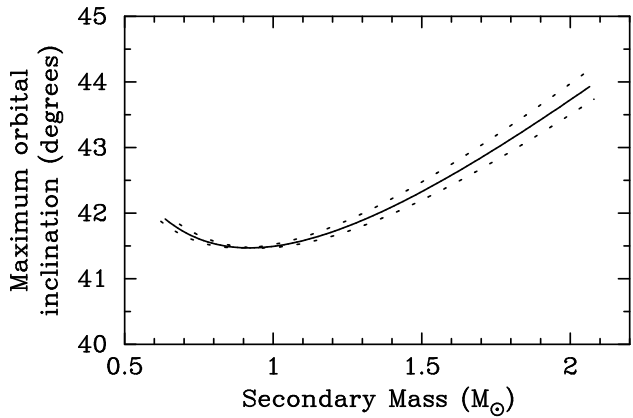}
\caption[]{The maximum inclination of the binary orbit of HD 209295,
consistent with the absence of ellipsoidal variability, as a function of
secondary mass. The full line corresponds to a primary mass of 1.84 M$_{\sun}$,
whereas the dotted lines delineate our 1$\sigma$ error estimate of 
$\pm$ 0.07 M$_{\sun}$.}
\end{figure}

Combining this information with the dependence of secondary mass on
inclination from the mass function of the binary (upper panel of Fig.~8),
we can derive a refined lower limit of the secondary mass. The result is
$M_2>1.04 M_{\sun}$. This lower limit on the secondary mass has two
implications. Firstly, it means that the primary is quite well within the
Roche limit. Secondly, the secondary mass is very close to the upper limit
$1.05 M_{\sun}$ derived in Sect. 3.2.1 assuming a main-sequence secondary.
In addition, the orbit of the system is also quite eccentric, which is
surprising as one would have expected circularization to have taken place.

Claret, Gim\'enez \& Cunha (1995) and Claret \& Cunha (1997) compared
theoretical orbital circularization and synchronization times for
main-sequence binaries with observations. From their results and using
the surface gravity of HD 209295 as an age indicator, we conclude that the 
orbital eccentricity of HD 209295 is inconsistent with normal main-sequence 
binary evolution. For a star similar in age to HD 209295 (log $t \approx 
8.86$ yr as inferred from comparison with pulsational and evolutionary
models), one would expect $e<0.1$. There is reason to believe, therefore,
that the secondary star may be a degenerate object. A white dwarf may have a 
mass in excess of 1.04\,$M_{\sun}$. However, in such a scenario orbital 
circularization would again have taken place. This is also expected because 
such a system must have undergone a previous mass transfer phase.

We investigated the ultraviolet fluxes of the object as measured by the TD-1 
satellite (Thompson et al. 1978) and compared it to the optical $uby$ fluxes. 
For this purpose, we converted the measured $uby$ magnitudes of HD 209295 to 
fluxes using the formulae of Gray (1998). We omitted the $v$ band because
it is dominated by the $H_\delta$ line. A Kurucz model atmosphere with 
$T_{\rm eff}=7750$~K (cf. Sect. 1.1) and log $g$=4.3 gave the best match.
We then compared the predicted UV fluxes from this fit with the TD-1 
measurements (Fig. 16). There is quite a strong UV excess in the TD-1 
measurements at 1965 and 2365\AA~compared to the model atmosphere, but little 
flux at 1565\AA. The shape of this excess resembles an energy distribution of 
an object of $T_{\rm eff}\approx$ 15000~K. This is a suggestion that a white 
dwarf companion could be responsible.

\begin{figure}
\includegraphics[width=110mm,viewport=-5 00 337 175]{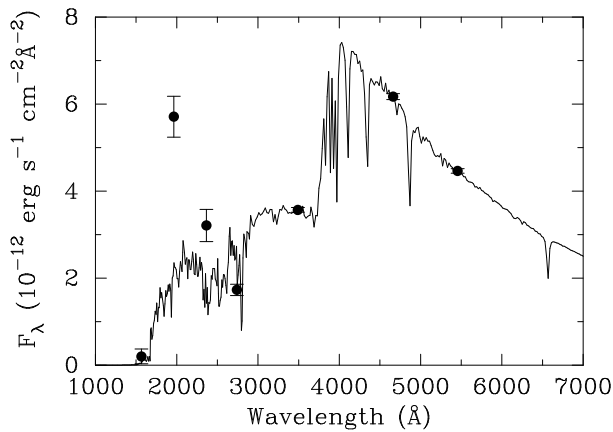}
\caption[]{Comparison of optical $uby$ and ultraviolet fluxes of HD
209295 to a $T_{\rm eff}=7750$~K, log $g$=4.3 Kurucz model atmosphere. A
clear UV excess is notable.}
\end{figure}

Consequently, we attempted to fit some white dwarf star model atmospheres
(Koester, private communication) for log $g$=8 and 10000~K $<T_{\rm eff}<$
16000~K to this excess. The $V$ flux was normalized to the value predicted
for a star at the distance of HD 209295. While we could reproduce the
approximate shape of the UV excess, the resulting white dwarf luminosity
is far too small. A white dwarf star would only be able to generate 0.1 --
0.2 \% of the observed UV excess. Hotter white dwarf models would be able
to explain a larger fraction of the excess, but they would be inconsistent 
with the low TD-1 flux at 1565\AA.

It is hard to imagine an astronomical object or a physical process which
could be made responsible for this UV excess and still be consistent with
other observations. A less massive white dwarf is inconsistent with our
mass constraints. A subdwarf would be seen in the optical and is
inconsistent with the primary being a Pop. I star. A late-type active
companion would be detected in X-rays. We need to leave the reason for the
UV excess unanswered; it may be possible that the TD-1 measurements are in
error. Unfortunately, the star was not observed by IUE (NASA-ESA 1999).

We also need to consider that the secondary may be a neutron star. In this
case, the high orbital eccentricity would be a result of the supernova
outburst, and the orbit would not be expected to circularize during the
main sequence life span of the present primary (see Zahn 1977, whose work
also suggests that the primary's rotation should not be synchronized with
the orbital motion of the compact secondary).

Comparing the orbital eccentricity and separation of the HD 209295 system
with the X-ray binary population simulations of Terman, Taam \& Savage
(1996) shows that the observed orbital parameters can be comfortably
explained in this scenario. The undetected companion of HD 209295 could
therefore be a neutron star, and the system will evolve into an
intermediate-mass X-ray binary after the present primary has left the main
sequence. The neutron star hypothesis is testable. The previous
evolutionary phases should have left their marks in the chemical
composition of HD 209295. However, the Str\"omgren metallicity index
$\delta m_1 = 0.016$ and our spectra appear normal.

If the unknown companion of HD 209295 were a neutron star, one would also
expect that it would be detectable in X-rays. However, the star has not been 
detected by ROSAT (Voges et al. 1999, 2000) or Einstein (Moran et al. 1996).
Finally, the neutron star should have suffered an impulse from the supernova
explosion, which might modify the space motion of the binary system. We
therefore calculated the galactic $U, V, W$ velocities of HD 209295 from
its $\gamma$ velocity from Table 7, its {\it Hipparcos} parallax (see
Sect. 1.1) and proper motion ($\mu_{\alpha}\cos \delta=26.84 \pm
0.59$~mas, $\mu_{\delta}= -58.82 \pm 0.41$~mas as measured by the {\it
Hipparcos} satellite (ESA 1997)). We find $U=-21$ km/s, $V=-12$ km/s and
$W=+28$ km/s (already corrected for the solar motion)
which is not unusual for an early-F main-sequence star (cf. Gilmore \&
Zeilik 2000). The neutron star interpretation for the secondary in the HD
209295 system therefore also has its weaknesses; we cannot at present
identify this star's nature with certainty.

\subsection{Evidence for forced oscillations}

In Sect.~3.4 we reported the discovery that many of the photometric
frequencies are exact integer multiples of the orbital frequency. This
raises the suspicion that they might not be free oscillation modes of the
star, but are rather triggered by tidal interactions. Tidally induced
nonradial oscillations have been searched for observationally, but the
results have not been very convincing, with the possible exception of the
slowly pulsating B star HD 177863 (De Cat 2001, Willems \& Aerts 2001). As
there are quite a number of frequencies which are exact integer multiples
of the orbital frequency, there is reason to believe that HD 209295 is the
best case for forced oscillations. Further support for this interpretation
needs to be sought. Some general information about tidal excitation is
therefore useful.

This effect has been studied by several authors theoretically, most often
in connection with neutron star/main sequence binaries. For instance,
Kumar, Ao \& Quatert (1995) showed that stellar p, f and low-order g-modes
are not easily excitable through tidal effects, but that intermediate-order 
g-modes may be excited. These are, of course, exactly the modes in which 
$\gamma$~Doradus stars (and SPB stars) pulsate.

Furthermore, tidally excited modes will have a shape adjusted to the
gravitational potential of the exciting star. The tidal deformation can be
decomposed into a linear superposition of the star's g- and p-mode
oscillations (Press \& Teukolsky 1977). Their dominating components are
the $\ell=2$, $|m|=2$ modes (e.g. see Kosovichev \& Severnyj 1983).
Unfortunately, we did not succeed in providing mode identifications for
one or more of the variations suspected to be forced oscillations.  
However, if we are really dealing with forced oscillations, they should be
approximately in phase at periastron where the tidal force is at maximum.

We have already listed pulsational phases of the different modes at a
periastron passage in Table 9. The phase values are similar and seem to
show some alignment of the supected tides. This is also the case for
$f_8=5f_{\rm orb}$ assuming it is a $\ell=2$, $|m|=2$ mode, as there is a
180\degr\, phase ambiguity. The variations are, however, not perfectly in 
phase, possibly due to nonadiabatic effects.

\subsection{The photometric and radial velocity amplitudes}

It is somewhat puzzling that there is little correspondence in amplitude
between the photometric and radial velocity data. The two highest-amplitude 
$\gamma$~Doradus modes in the photometry are undetected in the radial 
velocities. While the range in frequency of $\delta$~Scuti instability
is in approximate agreement in the photometry and radial velocities, there
is no clear match in frequencies between the two data sets.

Breger, Hutchins \& Kuhi (1976) determined the ratios of velocity to light
amplitudes in several $\delta$~Scuti stars and found values between
$2K/\Delta V$ of 50 -- 125 km\,s$^{-1}$\,mag$^{-1}$. Using this value for
the $\delta$~Scuti pulsations in HD 209295, we would expect the 
photometrically detected modes to have radial velocity amplitudes smaller
than 0.2 km/s, and hence they would not be detected. However, we detected
radial velocity variations with amplitudes of around 2 km/s in the
$\delta$~Scuti range, which should then generate photometric $V$
amplitudes of at least 15 mmag, but there is no trace of them. The only
plausible explanation is that the short-period variations in the
spectroscopy are not due to pulsation.

We can get limits to the radial velocity amplitude of the two independent
$\gamma$~Doradus modes of HD 209295 by fitting the radial velocities with
the photometric periods. This results in an upper limit of 1 km/s for
the amplitude of both modes. We then obtain 2$K/\Delta V < 26$
km\,s$^{-1}$\,mag$^{-1}$ for $f_1$ and 2$K/\Delta V < 35$
km\,s$^{-1}$\,mag$^{-1}$ for $f_2$.

In Table 10 we show all available amplitude determinations from photometric
and radial velocity measurements of $\gamma$~Doradus stars from the
literature. We only considered data sets in which simultaneous radial
velocity and photometry was obtained to avoid being sensitive to amplitude 
variations. Only those cases where individual mode periods could be
resolved are listed.

\begin{table}
\caption{A comparison of simultaneous $V$ filter light and radial velocity
amplitudes for $\gamma$~Doradus stars from the literature.}
\begin{center}
\begin{tabular}{cccccc}
\hline
Star & 2$K$ & $\Delta V$ & 2$K/\Delta V$ & $\ell$ & Ref.\\
 & (km/s) & (mmag) & (km\,s$^{-1}$\,mag$^{-1}$) & & \\
\hline
$\gamma$~Dor & 0.6 $\pm$ 0.2 & 23 $\pm$ 1 & 26 $\pm$ 9 & 3 & 1 \\
 & 2.6 $\pm$ 0.2 & 27 $\pm$ 1 & 96 $\pm$ 8 & 1 & \\
 & 1.2 $\pm$ 0.2 & 13 $\pm$ 1 & 92 $\pm$ 8 & 1 & \\ 
9 Aur & 1.7 $\pm$ 1.2 & 35 $\pm$ 1 & 48 $\pm$ 34 & 3 & 2, 3 \\
 & 3.0 $\pm$ 0.8 & 20 $\pm$ 1 & 148 $\pm$ 39 & 3 & 2, 3\\
 & 3.1 $\pm$ 1.1 & 18 $\pm$ 1 & 172 $\pm$ 61 & ? & 2, 3\\
HR 8330 & 5.2 $\pm$ 0.1 & 15 $\pm$ 1 & 350 $\pm$ 20 & 2 & 4, 5 \\
HD 68192 & 2.2 $\pm$ 0.2 & 21.7 $\pm$ 0.2 & 101 $\pm$ 10 & ? & 6\\
HD 209295 & $<1$ & 38.9 $\pm$ 0.2 & $< 26$ & 1 & 7\\
	 & $<1$ & 28.7 $\pm$ 0.2 & $< 35$ & 1 & 7\\
\hline
\end{tabular}
\end{center}
References: 1: Balona et al. (1996), 2: Zerbi et al. (1997), 3: Aerts \&
Krisciunas (1996), 4: Kaye et al. (1999b), 5: Aerts \& Kaye (2001), 6:
Kaye et al. (1999c), 7: this paper
\end{table}

It appears that there is an order of magnitude spread in the 2$K/\Delta
V$ values for the different stars, but also for different modes of the
same star. The limits we obtained for HD 209295 are at the lower end of
the range to be found in Table 1. However, the lowest 2$K/\Delta V$ values
were obtained for modes of $\ell=3$, not for $\ell=1$ as suggested for
the two independent modes of HD 209295.

Aerts \& Krisciunas (1996) offered an explanation for such diverse
behaviour: the photometric variations are mostly due to temperature
variations, which hardly affect the radial velocities. In addition Aerts
\& Krisciunas (1996) suggest that, due to stellar rotation, toroidal
corrections become important. These may be the cause for the large scatter
in 2$K/\Delta V$ in Table 10. High-resolution, high signal-to-noise
spectra of the more rapidly rotating $\gamma$~Dor stars with simultaneous
photometry are necessary to test this hypothesis.

\section{Theory}

We have carried out a stability analysis of the pulsations to check
whether the inferred evolutionary state of HD 209295 is consistent with its 
pulsational behaviour, and whether tidally induced oscillations are 
reasonable.

\subsection{Stability analysis}

We used the Warsaw-New Jersey stellar evolution and the NADROT pulsation
codes to generate a series of models for HD 209295. We then investigated
the pulsation modes for stability. The results for the model which best
matches the parameters of HD 209295 ($M=1.8M_{\sun}$, log $T_{\rm eff}$ =
3.88, log $L$ = 1.10, log $g$=4.03) are shown in Fig. 17.

\begin{figure}
\includegraphics[width=107mm,viewport=-5 00 335 320]{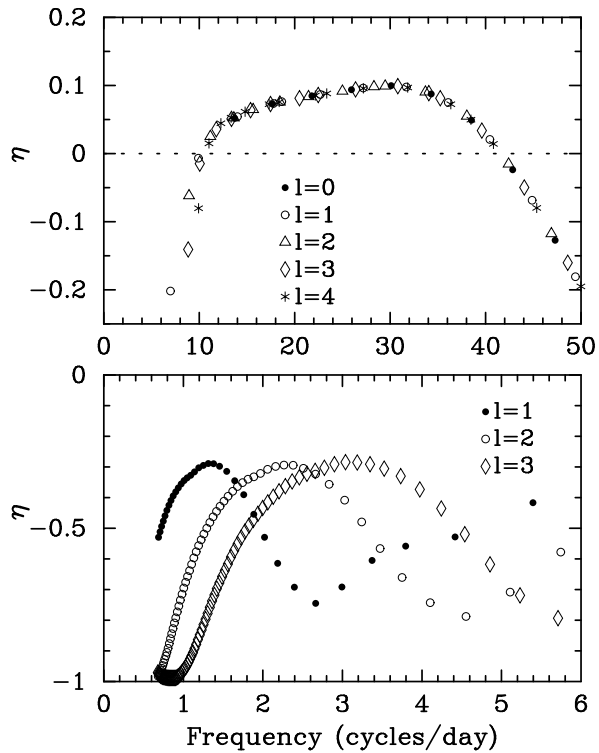}
\caption[]{Normalized linear growth rates, $\eta$, for a model matching the 
effective temperature and surface gravity HD 209295. $\eta=1$ means
that a mode is driven throughout the whole model, whereas $\eta=-1$ means
that a mode is damped throughout the whole model; modes with $\eta>0$ are
pulsationally unstable. Upper panel: the whole computed frequency range.
Lower panel: the $\gamma$~Dor pulsation frequency domain. In certain frequency
regions (depending on $\ell$), less damping is present.}
\end{figure} 

The observed frequency domain of the $\delta$~Scuti pulsations of HD
209295 is very well reproduced. The unstable frequency range does not
depend on $\ell$, which is also consistent with the observations. $\gamma$
Doradus modes are not driven in our models, as this currently requires a
special treatment of convection (Guzik et al. 2000). In our computations
we used the standard mixing-length convection theory and we ignored the 
Lagrangian perturbation of the convective energy flux when computing 
oscillations. Our attempts to reproduce the results by Guzik et al. (2000) 
concerning pulsational instability were unsuccessful, and we plan to study 
this matter in the future.

It is interesting to note that there are frequency regions in which
damping is not as strong, and that they depend on $\ell$. We performed
these computations up to $\ell=8$, and found that the trend is maintained.
The $\ell=1$ and $\ell=2$ frequency regions coincide with the ones
actually observed.

\subsection{Tidal excitation}

We investigated the possibility of tidally excited oscillations in
HD\,209295 by determining the amplitude of tidally induced radial-velocity
variations using the expressions derived by Willems \& Aerts (2001). We
calculated both the free and the forced oscillations of a $1.8\,M_\odot$
stellar model in the linear, adiabatic approximation. Since the amplitude
may become quite large in the adiabatic approximation, we restricted
ourselves to calculating the radial-velocity variations at orbital periods
for which the relative differences between the forcing frequencies of the
dynamic tides and the eigenfrequencies of the free oscillation modes are
not too small (for details see Willems \& Aerts 2001).

The resulting amplitudes of the tidally induced radial-velocity variations
seen by an observer are displayed in Fig.~18 as a function of the orbital
period. The companion mass is assumed to be $1.5\,M_\odot$ which
corresponds to an orbital inclination of 31.2$\degr$ and to a rotational
frequency between $f_{\rm rot} = 1.635 $\,d$^{-1}$ and $f_{\rm rot} =
1.998 $\,d$^{-1}$.

\begin{figure}
\includegraphics[width=115mm,viewport=20 20 762 565]{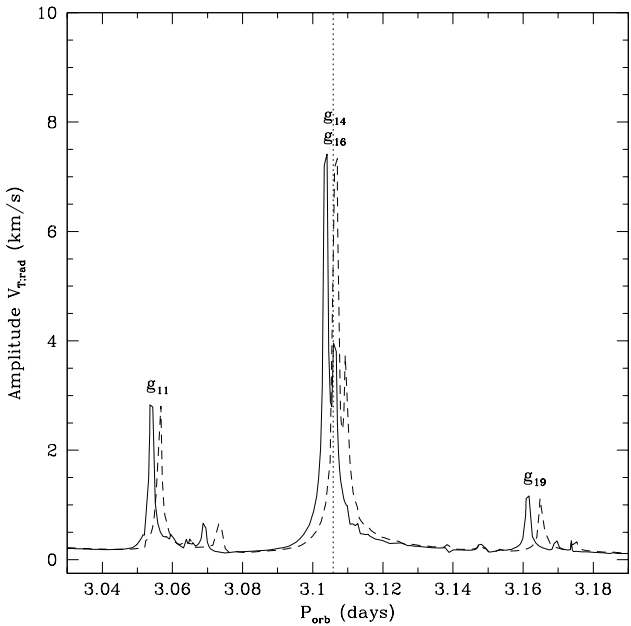}
\caption[]{Observed amplitude of the tidally induced radial velocity
variations in a model for HD\,209295. The full line corresponds to a
rotational frequency $f_{\rm rot}=1.852 $\,d$^{-1}$ and the dashed line
to a rotational frequency $f_{\rm rot}=1.854 $\,d$^{-1}$. The orbital
period of HD\,209295 is indicated by the dotted vertical line.}
\end{figure}

A high peak in the amplitudes is seen to occur near the observed orbital
period of HD\,209295. The peak results from the resonant excitation of the
modes $g_{14}^+$ and $g_{16}^+$ by the dynamic tides associated with the
forcing frequencies $3\,f_{\rm orb}$ and $4\,f_{\rm orb}$, respectively.
Both tides are of spherical degree $\ell=2$ and azimuthal order $m=-2$.
This result is quite encouraging, as a signal at $3\,f_{\rm orb}$ was
detected in the photometry, and one at $4\,f_{\rm orb}$ was suspected. On
the other hand, not all of the detected orbital harmonics are excited in
the model, but they can be reconciled with linear combinations of these
two signals.

We need to make several remarks here. Firstly, the above ``solution'' is
probably not unique. In addition, the calculations have been performed
using an adiabatic code, so they show the presence of the resonances, but
not the real amplitudes. Secondly, attempts with lower companion masses
(and thus lower rotational frequencies) were not very successful. This is
interesting, as it may help to constrain the nature of the orbital
companion of HD 209295.

We conclude that resonant excitation is possible, but for precise results
there are still too many degrees of freedom. The prospect of developing
tides as a probing tool is a promise for future studies.

\section{Conclusions and outlook}

By means of 128 h of time-resolved high-resolution spectroscopic and 280 h
of time-series photometric observations we have shown that HD 209295 is
not only the first pulsating star to be a member of two classes (a
$\gamma$~Doradus as well as a $\delta$~Scuti star), but we have also
discovered good evidence for the presence of tidally excited pulsation
modes. The binary companion is probably a neutron star or a white dwarf.

These findings make HD 209295 a rather interesting astrophysical laboratory. 
We were able to confront photometric and spectroscopic mode identification 
methods. Whereas the method by Telting \& Schrijvers (1997) yielded reasonable 
assignments for the two independent modes dominating the photometry (both are 
$\ell=1, |m|=1$), analysing the photometric colour amplitude ratios and phase
differences did not yield any meaningful results. The necessity of using as 
many mode identification methods as possible to secure reliable results is
stressed.

We showed that the presence of tidally induced modes can be explained by 
theoretical models, but not all the observations can be understood. It would 
be important to calculate more realistic models of tidal interaction.

We do not fully understand the wealth of information provided by the
variations in HD 209295. Large coordinated efforts will be required for
further progress as our data are already fairly extensive. Another multi-site 
photometric and spectroscopic campaign on an even larger scale would be 
desirable. This should be carried over a long time base in order to resolve 
the $\gamma$~Dor pulsations, and a large data set is needed to detect more 
$\delta$~Scuti modes. Large telescopes are required for the spectroscopy to 
obtain high time resolution and a high signal to noise ratio. This is
necessary for identifying further modes and for detailed line-profile
analysis. Photometric observations in more filters than previously obtained
are necessary, perhaps extending the wavelength range into the near infrared. 
UV and infrared spectroscopy would be useful in a search for the companion
star, and an abundance analysis may reveal an unusual chemical composition
caused by the possible previous evolution of the system through mass
transfer, common envelope and/or supernova stages.

\section*{ACKNOWLEDGEMENTS}

We are indebted to Lisa Crause for obtaining and reducing the infrared
photometry of HD 209295 at our request and for commenting on a draft
version of this paper. DJJ and BW thank the British Particle Physics and
Astronomy Research Council (PPARC) for post-doctoral research fellowships.
DJJ would also like to acknowledge the continued positive influences of
Mrs J Pryer, for much love and support during this work, and the Royal
Society for a European research grant. ERC thanks S. Potter, D. Romero and
E. Colmenero for their support. AAP would like to thank L. A. Balona and
the staff of the SAAO (Cape Town) for their hospitality during his visit
to the institute. BW is grateful to Antonio Claret for providing a 1.8
$M_{\sun}$ stellar model. GH thanks Darragh O'Donoghue for giving him one
night of his observing time for this project and Detlev Koester for
providing white dwarf model atmospheres. GH also wishes to express his
gratitude to several colleagues for helpful comments: Brian Warner, Marten
van Kerkwijk, Matt Burleigh and Stephen Potter. Ennio Poretti and Kevin
Krisciunas are thanked for commenting on a draft version of this paper.

\end{document}